\documentclass[useAMS,usenatbib]{mn2e}
\usepackage{astroshortcuts} 
\usepackage{graphicx}
\usepackage{bm}
\usepackage{ifthen}
\newboolean{useepsfigures}
\setboolean{useepsfigures}{false}

\title{The Graininess of Dark Matter Haloes}

\author[M. Zemp et al.]{Marcel Zemp$^{1,2}$\thanks{mzemp@umich.edu}, 
J\"{u}rg Diemand$^{2,6}$, 
Michael Kuhlen$^3$,
Piero Madau$^2$,\newauthor
Ben Moore$^4$,
Doug Potter$^4$, 
Joachim Stadel$^4$ and
Lawrence Widrow$^5$\\
$^1$Department of Astronomy, University of Michigan, 500 Church Street, Ann Arbor, MI 48109\\
$^2$Department of Astronomy \& Astrophysics, University of California Santa Cruz, 1156 High Street, Santa Cruz, CA 95064\\
$^3$School of Natural Sciences, Institute for Advanced Study, Einstein Lane, Princeton, NJ 08540\\
$^4$Institute for Theoretical Physics, University Zurich, Winterthurerstrasse 190, 8057 Zurich\\
$^5$Department of Physics, Engineering Physics, and Astronomy, Queen's University, Kingston, ON K7L 3N6\\
$^6$Hubble Fellow}

\begin{document}

\pagerange{\pageref{firstpage}--\pageref{lastpage}} \pubyear{2008}

\maketitle

\label{firstpage}

\begin{abstract}
We use the recently completed one billion particle {\it Via Lactea II} $\Lambda$CDM simulation to investigate local properties like density, mean velocity, velocity dispersion, anisotropy, orientation and shape of the velocity dispersion ellipsoid, as well as  structure in velocity space of dark matter haloes. We show that at the same radial distance from the halo centre, these properties can deviate by orders of magnitude from the canonical, spherically averaged values, a variation that can only be partly explained by triaxiality and the presence of subhaloes. The mass density appears smooth in the central relaxed regions but spans four orders of magnitude in the outskirts, both because of the presence of subhaloes as well as of underdense regions and holes in the matter distribution. In the inner regions the local velocity dispersion ellipsoid is aligned with the shape ellipsoid of the halo. This is not true in the outer parts where the orientation becomes more isotropic. The clumpy structure in local velocity space of the outer halo can not be well described by a smooth multivariate normal distribution. {\it Via Lactea II} also shows the presence of cold streams made visible by their high 6D phase space density. Generally, the structure of dark matter haloes shows a high degree of graininess in phase space that cannot be described by a smooth distribution function. 
\end{abstract}


\section{Introduction}

Spherical averaging is a commonly used method to describe the characteristics of dark matter haloes that form by gravitational collapse in a cosmological environment. For example one of the basic characteristics of a dark matter halo is its spherically-averaged density profile which can be well described by a smooth function within resolved scales \citep[e.g.][]{1996ApJ...462..563N,1998ApJ...499L...5M,2005MNRAS.364..665D,2007ApJ...657..262D,2008Natur.454..735D,2008arXiv0808.2981S}. Other examples include the velocity dispersion and anisotropy profiles.

Consider a set of observers at a given distance from the halo centre and capable of measuring local properties of the halo (e.g., density, velocity dispersion, velocity anisotropy). In a smooth, spherically symmetric halo, these measurements would yield identical values, up to statistical fluctuations. However, it is well known that dark matter haloes that form in pure dark matter simulations, which neglect possible baryonic effects in the centre, are in general triaxial: close to prolate in the central part and becoming rounder in the outskirts of the halo \citep[e.g.][]{1991ApJ...368..325K, 1991ApJ...378..496D, 2006MNRAS.367.1781A, 2007MNRAS.376..215B, 2007ApJ...671.1135K}. Such a shape variation obviously leads to a significant variance in local properties compared to the spherically averaged value at a given radius \citep{2006PASA...23..125K}. In addition, haloes have a high level of subhaloes which also effect results of local measurements. But local deviations from a smooth model go beyond these shape and subhalo driven variations, as we demonstrate in this paper. For example, overdensities are expected due to the presence of subhaloes, but we also find underdense regions, which are unexpected in a smooth triaxial background halo with subhaloes (see section \ref{sec:den}).

One aim of this paper is to quantify the degree of variation between local and spherically averaged values for key quantities such as the density and velocity dispersion. A better understanding of the local variations of these properties is essential in order to obtain a better description of the phase space structure of dark matter haloes and their formation process. 

Such local variations can have profound consequences. For example, in dynamical models of dark matter haloes and galaxies it is often assumed that most properties are just a function of radius \citep[e.g.][]{2005MNRAS.363.1057D,2007A&A...471..419B,2008arXiv0809.0901V}. From the work presented here it becomes clear that this is not a valid assumption for the anisotropy (see section \ref{sec:ani}) and a lot of information is smeared out by spherically averaging. Models based on simple analytical forms of the distribution function, which are also used for generating $N$-body realisations of dark matter haloes and galaxies \citep[e.g.][]{2004ApJ...601...37K,2008MNRAS.386.1543Z,2008MNRAS.387.1719Z}, often have the property that the local bulk motion is zero and the velocity distribution function is similar to a smooth multivariate normal distribution (if it is not even deliberately set to a multivariate normal distribution as in \cite{1993ApJS...86..389H}). Both of these characteristics are not valid in the outskirts of dark matter haloes that form in a hierarchical cosmological context (see sections \ref{sec:meanvel} and \ref{sec:velspace}). As we see, most of these effects are not taken into account in current dynamical models of dark matter haloes and galaxies and they will be a challenge to be modelled with more realistic distribution functions. 

In addition, it is also important to quantify these variations locally at the Earth's position within the Galaxy in order to interpret results from present and future direct dark matter detection experiments on Earth
\citep[e.g.][]{2001PhRvD..64f3508M,2001PhRvD..64h3516S,2002PhRvD..66f3502H,2003PhRvL..90u1301S,2008PhRvD..77j3509K,2008MNRAS.385..236V,2008arXiv0808.0704F}. While small subhalos contribute significantly to the indirect detection signal \citep[e.g.][]{2006ApJ...649....1D,2008ApJ...686..262K} and infall caustics have no significant effect \citep{2008ApJ...680L..25D}, it is worthwhile also to investigate whether additional lumpiness from tidal debris might enhance the annihilation rate (see section \ref{sec:den})\footnote{After submitting this work an independent, analytic study of this question \citep{2008arXiv0811.1582A} has appeared. Our simulation results agree that such an enhancement is rather small (see section \ref{sec:den}).}. 

Also recent results from surveys like SDSS and RAVE need a better understanding of the structure of the outer halo. There is a lot of known structure in the Galactic stellar halo like for example the Sagittarius stream, the Monoceros stream or the Virgo overdensity \citep[e.g.][]{2007ApJ...660.1264M,2008ApJ...683..750C,2008AJ....135.2013C}. Obviously, these features are baryonic, but stars are collisionless too, and it's probable that these features originate from infalling dark matter dominated objects.

In general, a more detailed picture of the phase space structure of dark matter haloes is needed in order to understand and model their properties that are set by the hierarchical formation process. This work is a further important step towards that aim.

Here we present a study of the distribution of these properties as a function of galactocentric distance of a Milky Way size dark matter halo with data from the Via Lactea II project \citep{2008Natur.454..735D}. In section \ref{sec:data} we give a summary of the simulation data used in this study and present the results in section \ref{sec:properties}. We then discuss our results and present the conclusions in section \ref{sec:conclusions}. In appendix \ref{sec:comparison} we give a comparison of the results with lower resolution simulations and different definitions of locality. 

\section{Simulation and data}\label{sec:data}

\begin{table*}
	\caption{Summary of shell properties.}
	\label{tab:shellsummary}	
	\begin{tabular}{llcccccc}
		\hline
		$r$ & $[\kpc]$ & 8 & 25 & 50 & 100 & 200 & 400\\
		\hline
	
		$\rho$ & $[\Mo~\pc^{-3}]$ & $1.056 \times 10^{-2}$ & $1.304 \times 10^{-3}$ & $3.148 \times 10^{-4}$ & $5.243 \times 10^{-5}$ & $7.020 \times 10^{-6}$ & $1.079 \times 10^{-6}$\\
		$r_{\mathrm{sph}}$ & $[\kpc]$ & 0.5 & 1.004 & 1.612 & 2.930 & 5.728 & 10.69\\
	
		$\left\langle M_{\mathrm{sph}}\right\rangle$ & $[\Mo]$ & $5.555 \times 10^{6}$ & $5.600 \times 10^{6}$ & $5.557 \times 10^{6}$ & $5.791 \times 10^{6}$ & $5.566 \times 10^{6}$ & $6.014 \times 10^{6}$\\
	
		$\tilde{\rho}$ & $[\Mo~\pc^{-3}]$ & $1.059 \times 10^{-2}$ & $1.305 \times 10^{-3}$ & $3.140 \times 10^{-4}$ & $5.364 \times 10^{-5}$ & $7.084 \times 10^{-6}$ & $1.094 \times 10^{-6}$\\
	
		$\tilde{v}_{\mathrm{r}}$ & $[\km~\s^{-1}]$ & -0.2116 & -1.075 & -0.2053 & -4.207 & -2.975 & 9.285\\
	
		$\tilde{v}_\varphi$ & $[\km~\s^{-1}]$ & 0.7794 & 2.946 & -8.476 & -0.1219 & 0.6685 & 9.969\\
	
		$\tilde{v}_\vartheta$ & $[\km~\s^{-1}]$ & -0.3097 & -0.8743 & 0.7920 & -4.188 & 4.331 & -12.52\\
	
		$\tilde{\sigma}_{\mathrm{tot}}$ & $[\km~\s^{-1}]$ & 239.5 & 246.2 & 219.1 & 187.8 & 147.5 & 122.2\\
	
		$\tilde{\sigma}_{\mathrm{r}}$ & $[\km~\s^{-1}]$ & 144.1 & 158.5 & 145.3 & 125.9 & 98.99 & 73.03\\
	
		$\tilde{\sigma}_{\mathrm{\varphi}}$ & $[\km~\s^{-1}]$ & 143.1 & 145.4 & 129.4 & 109.8 & 83.36 & 76.23\\
	
		$\tilde{\sigma}_{\mathrm{\vartheta}}$ & $[\km~\s^{-1}]$ & 127.0 & 119.9 & 100.9 & 85.82 & 70.86 & 61.62\\
	
		$\tilde{\beta}$ & [1] & 0.1188 & 0.2938 & 0.3620 & 0.3872 & 0.3893 & 0.09919 \\
	
		$\tilde{\alpha}_a$ & [$^\circ$] & 0.1480 & 4.633 & 3.164 & 4.893 & 16.18 & 37.31\\
		
		$\tilde{\alpha}_b$ & [$^\circ$] & 1.121 & 5.596 & 3.577 & 13.06 & 16.53 & 48.12\\
	
		$\tilde{\alpha}_c$ & [$^\circ$] & 1.117 & 3.166 &  3.532 & 12.80 & 5.159 & 31.24\\
	
		$\tilde{T}$ & [1] & 0.6308 & 0.4893 & 0.5066 & 0.3894 & 0.6017 & 0.6091\\
	
		$\sigma(\bar{\rho})$ & $[\Mo~\pc^{-3}]$ & $2.746 \times 10^{-3}$ & $6.748 \times 10^{-4}$ & $2.005 \times 10^{-4}$ & $7.436 \times 10^{-5}$ & $1.603 \times 10^{-5}$ & $5.748 \times 10^{-6}$\\
	
		$\sigma(\bar{v}_{\mathrm{r}})$ & $[\km~\s^{-1}]$ & 5.681 & 14.03 & 16.71 & 23.24 & 26.47 & 30.57\\
	
		$\sigma(\bar{v}_{\varphi})$ & $[\km~\s^{-1}]$ & 6.077 & 11.45 & 19.44 & 28.22 & 27.08 & 36.59\\
	
		$\sigma(\bar{v}_{\vartheta})$ & $[\km~\s^{-1}]$ & 4.474 & 11.54 & 13.96 & 19.52 & 20.91 & 29.13 \\
	
		$\sigma(\bar{\sigma}_{\mathrm{tot}})$ & $[\km~\s^{-1}]$ & 6.745 & 9.491 & 8.780 & 12.27 & 15.60 & 21.20\\
	
		$\sigma(\bar{\sigma}_{\mathrm{r}})$ & $[\km~\s^{-1}]$ & 7.013 & 6.405 & 7.724 & 11.22 & 13.60 & 18.72\\
	
		$\sigma(\bar{\sigma}_{\varphi})$ & $[\km~\s^{-1}]$ & 10.77 & 11.04 & 12.23 & 13.84 & 14.71 & 20.81\\
	
		$\sigma(\bar{\sigma}_{\vartheta})$ & $[\km~\s^{-1}]$ & 11.42 & 16.21 & 15.17 & 14.75 & 16.86 & 17.79\\
	
		$\sigma(\bar{\beta})$ & [1] & 0.2059 & 0.1509 & 0.1165 & 0.1863 & 0.2534 & 0.8370\\
	
		$\sigma(\bar{\alpha}_a)$ & [$^\circ$] & 7.244 & 22.61 & 24.29 & 23.76 & 23.25 & 23.43\\
	
		$\sigma(\bar{\alpha}_b)$ & [$^\circ$] & 7.990 & 22.65 & 24.84 & 24.12 & 19.91 & 22.37\\
	
		$\sigma(\bar{\alpha}_c)$ & [$^\circ$] & 6.249 & 17.83 & 22.77 & 22.67 & 23.25 & 23.03\\
	
		$\sigma(\overline{T})$ & [1] & 0.1692 & 0.2208 & 0.2075 & 0.1931 & 0.2038 & 0.2036\\
	
		$f_{\mathrm{subhalo}}$ & [1] & 0.01060 & 0.05870 & 0.09630 & 0.1739 & 0.2492 & 0.2779\\
	
		$f_{\mathrm{empty}}$ & [1] & 0 & 0 & $5.000 \times 10^{-4}$ & $9.000 \times 10^{-4}$ & $1.500 \times 10^{-3}$ & $2.120 \times 10^{-2}$\\
	
		$G(\bar{\rho})$ & [1] & 0.1415 & 0.2813 & 0.3384 & 0.3761 & 0.3828 & 0.6193\\
			
		\hline
	\end{tabular}
	\medskip
	\begin{flushleft}
	Rows are: 
	galactocentric distance $r$, 
	spherically averaged density, 
	radius of spheres,
	ensemble averaged mass in spheres, 
	in shell spherically averaged value of density, 
	mean radial velocity, 
	mean $\varphi$-velocity, 
	mean $\vartheta$-velocity,
	total velocity dispersion,
	radial velocity dispersion, 
	$\varphi$-velocity dispersion, 
	$\vartheta$-velocity dispersion,
	anisotropy parameter,
	angle between long axis of shape ellipsoid and long axis of velocity ellipsoid,
	angle between intermediate axis of shape ellipsoid and intermediate axis of velocity ellipsoid,
	angle between short axis of shape ellipsoid and short axis of velocity ellipsoid and
	triaxiality parameter of the velocity dispersion ellipsoid.
	Further we have the ensemble dispersions of the above mentioned quantities.
	The last three rows are the fraction of spheres that are affected by subhaloes, the fraction of empty spheres and the Gini coefficient for the probability density functions of the local mean density.
	\end{flushleft}
\end{table*}

For the analysis presented here, we used data from the  Via Lactea II simulation \citep{2008Natur.454..735D} which simulates the assembly of a Milky Way size cold dark matter halo within a $\Lambda$CDM universe. The initial conditions at a starting redshift of $z=104.3$ consist of a 40 comoving Mpc periodic box and were generated with a modified, parallel version of GRAFIC2 \citep{2001ApJS..137....1B}. We use the traditional method of refining a region of interest with a large number of particles and leaving the rest on lower resolution so that we correctly account for the large scale tidal forces \citep{1993ApJ...412..455K, 2001ApJS..137....1B}. In total, the simulation consists of more than $10^9$ high resolution particles with a mass of $m_{\mathrm{p}} = 4098~\Mo$ and a gravitational softening length of 40 pc. We used the WMAP 3-year cosmological parameters \citep{2007ApJS..170..377S} with $\Omega_{\mathrm{M},0} = 0.238$, $\Omega_{\Lambda,0} = 0.762$ and $H_0 = 73~\km~\s^{-1}~\Mpc^{-1}$, $\sigma_8 = 0.74$ and $n_s = 0.951$. 

The time evolution until redshift $z=0$ was performed with the parallel tree-code PKDGRAV2 \citep{2001PhDT........21S}. PKDGRAV2 uses a fast multipole expansion technique in order to calculate the forces with hexadecapole precision and a time-stepping scheme that is based on the true dynamical time of the particles with an accuracy parameter of $\eta_{\mathrm{D}} = 0.06$ \citep{2007MNRAS.376..273Z, 2008Stadel}. The simulation used close to $10^6$ CPUh on the Jaguar Cray XT3 super-computer at the Oak Ridge National Laboratory\footnote{\texttt{http://www.ornl.gov}}. 

Throughout the paper, we only use data from the $z=0$ snapshot. We present all quantities in physical units and in a coordinate system centred on the particle with the deepest potential within the dark matter halo. A detailed discussion of the Via Lactea II simulation at $z=0$ can be found in \citet{2008Natur.454..735D} and we report here only the main characteristics. The radius $r_{\mathrm{200b}}$, where the enclosed density is 200 times the mean matter density (background), is $r_{\mathrm{200b}} = 402.1~\kpc$ and contains a total mass of $M_{\mathrm{200b}} = M(r_{\mathrm{200b}}) = 1.917 \times 10^{12}~\Mo$. The maximum circular velocity of the halo is $v_{\mathrm{cmax}} = 201.3~\km~\s^{-1}$ and is reached at the radius $r_{v_{\mathrm{cmax}}} = 59.83~\kpc$. We determine the shape of the halo at 20 kpc by diagonalising the shape tensor\footnote{Often the tensor $\mathbf{S}$ is incorrectly denoted as the inertia tensor. However, the tensor $\mathbf{S}$ does not give the correct relation between the angular momentum vector $\vec{L}$ and the angular velocity vector $\bm\omega$ which is given by $\vec{L} = \mathbf{I}~\bm\omega$ where $I_{ij} \equiv \sum_k m_k [r^2_k \delta_{ij} - (x_k)_i (x_k)_j]$ is the correct definition of the inertia tensor. Of course, the tensors $\mathbf{I}$ and $\mathbf{S}$ have the same eigenvectors since they differ only by a diagonal tensor with one eigenvalue of algebraic multiciply of 3. Though, the eigenvalues of $\mathbf{S}$ and $\mathbf{I}$ are obviously different and have entirely different meanings.} $\mathbf{S}$ with elements
\begin{equation}
S_{ij} \equiv \frac{\sum_k m_k (x_k)_i (x_k)_j}{\sum_k m_k}
\end{equation}
through an iterative procedure that adapts to the local shape as described in \citet{1991ApJ...368..325K} so that the summation over the particles only contains those within the triaxial ellipsoid. We then rotate the whole halo so that the long axis is the $x$-axis, the intermediate axis is the $y$-axis, and the short axis is the $z$-axis. The resulting axis ratios at 20 kpc are $b/a = 0.6154$ and $c/a = 0.5227$ where $a \geq b \geq c$ are the square roots of the eigenvalues of the shape tensor. Of course, a dark matter halo is not a rigid body and the axis ratios as well as their orientations can change with radius, with haloes typically becoming rounder in their outer part. The orientation of the principle axis is constant with radius (to within a few degrees) between 20 kpc and 400 kpc.

\section{Local Properties}\label{sec:properties}

\subsection{Procedure}

In order to measure local properties of the dark matter halo, we chose the following procedure. We consider six distinct distances: 8, 25, 50, 100, 200 and 400 kpc from the halo centre. At each distance we randomly sample $N_\mathrm{sample}=10^4$ spheres with a radius $r_\mathrm{sph}(r)$ depending only on the distance and chosen to include of $\mathcal{O}(10^3)$ particles. Specifically we set $r_\mathrm{sph}(8~\kpc) = 0.5~\kpc$ and calculate the radii at other distances by
\begin{equation}
r_\mathrm{sph}(r) = r_\mathrm{sph}(8~\kpc) \sqrt[3]{\frac{\rho(8~\kpc)}{\rho(r)}},
\end{equation}
where $\rho(r)$ is the measured spherically averaged density at radius $r$. The number of sampling spheres is large enough so that even at 400 kpc the whole sky, as seen from the centre of the galaxy, is covered more than once. A summary of the characteristics of the shells at different distances from the galactic centre is given in Table \ref{tab:shellsummary}. In order to save space, in most cases we only show the plots for 8, 25, 100 and 400 kpc in the following analysis.

In order to assess the influence of the triaxial shape of the halo, we also consider 3 sub-sets of spheres whose centres lie within a cone of 20$^\circ$ along the long ($x$), the intermediate ($y$) respectively the short ($z$) axis, i.e. spheres that fulfil the condition
\begin{equation}
|\vec{e}_{\mathrm{sph}} \cdot \vec{e}_k| > \cos(20^\circ) \quad \mathrm{with} \quad k = x, y ~\mathrm{or}~ z
\end{equation}
where $\vec{e}_{\mathrm{sph}}$ is the unit vector in galactocentric coordinates towards the centre of the sphere and $\vec{e}_k$ is the unit vector along the $x$-, $y$- respectively $z$-axis. It is expected that each sub-set contains approximately 6\% of the total number of spheres and the actual numbers lie within the statistical range. The slightly twisted shape of the dark matter halo has only a small influence on the sub-sets along the different axes at different distances $r$ since the deviations of the eigenvector directions are only a few degrees which is small compared to the cone opening angle of 20$^\circ$.

Furthermore we consider the sub-sample of spheres that are affected by subhaloes. We take a group catalogue of the Via Lactea II simulation that contains all subhaloes with a peak circular velocity greater than $2~\km~\s^{-1}$, of which there are 26182 within 500 kpc. These groups are identified by a friends-of-friends method in phase space (6DFOF) as described in \cite{2006ApJ...649....1D}. By a density profile fitting algorithm, we determine a size $r_{\mathrm{sh}}$ for each subhalo. A sphere with centre at location $\vec{x}_{\mathrm{sph}}$ is then deemed affected by a subhalo if the criterion
\begin{equation}
|\vec{x}_{\mathrm{sph}}-\vec{x}_{\mathrm{sh},k}| \leq r_{\mathrm{sph}} + r_{\mathrm{sh},k}
\end{equation}
is fulfilled for any subhalo $k$ with radius $r_{\rm sh,k}$ and centred at $\vec{x}_{\mathrm{sh},k}$, i.e. the subhalo $k$ at least partially overlaps the sphere. We give the fraction of spheres that are affected by subhaloes in Table \ref{tab:shellsummary}. This fraction increases with distance from the centre of the galaxy, reaching a value of approximately 28\% at 400 kpc. The fraction of subhalo-affected spheres is a lower limit since more subhaloes would survive in higher resolution simulations at all distances from the centre of the galaxy.

Throughout this paper we use the following notation scheme: quantities calculated by averaging over particles within a sphere are denoted by $\bar{q}$, 
quantities calculated by averaging over particles within a shell from $r - r_{\mathrm{sph}}(r)$ to $r + r_{\mathrm{sph}}(r)$ are denoted by $\tilde{q}$, and averages over the ensemble of $N_{\mathrm{sample}}$ spheres at radius $r$ are denoted by $\left\langle \bar{q} \right\rangle$, where we also might use indices in order to denote ensemble averages over only sub-sets of the spheres, for any given quantity $q$.

\subsection{Local densities}\label{sec:den}

\begin{figure*}
	\centering
	\ifthenelse{\boolean{useepsfigures}}{
	\includegraphics[width=0.495\textwidth]{shell8kpc_rho}
	\includegraphics[width=0.495\textwidth]{shell25kpc_rho}\\
	\includegraphics[width=0.495\textwidth]{shell100kpc_rho}
	\includegraphics[width=0.495\textwidth]{shell400kpc_rho}
	}{
	\includegraphics[width=0.495\textwidth,bb=0 0 574 574]{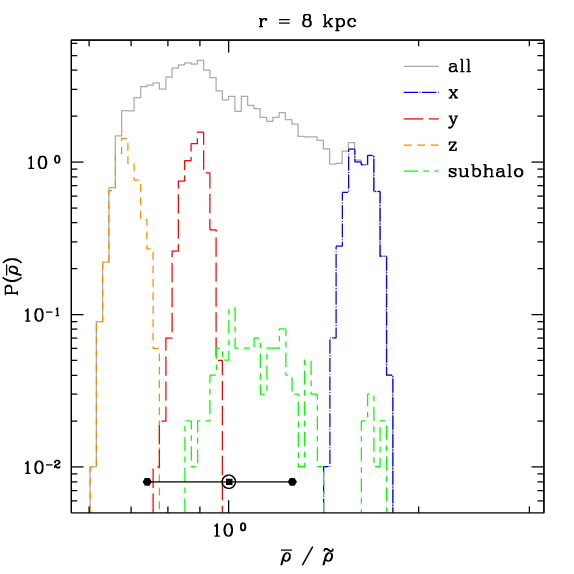}
	\includegraphics[width=0.495\textwidth,bb=0 0 574 574]{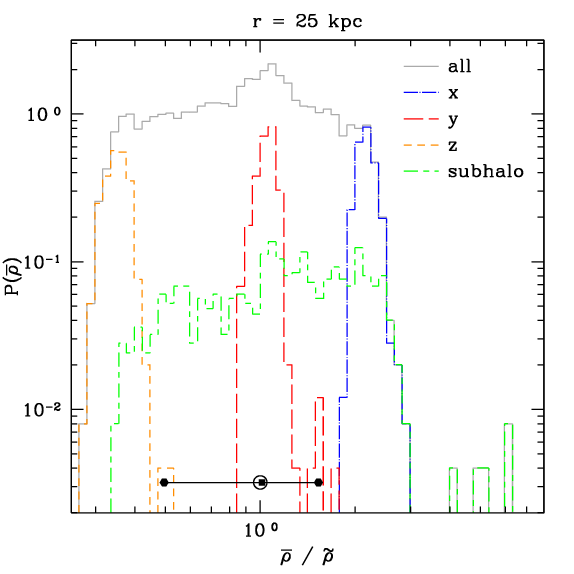}\\
	\includegraphics[width=0.495\textwidth,bb=0 0 574 574]{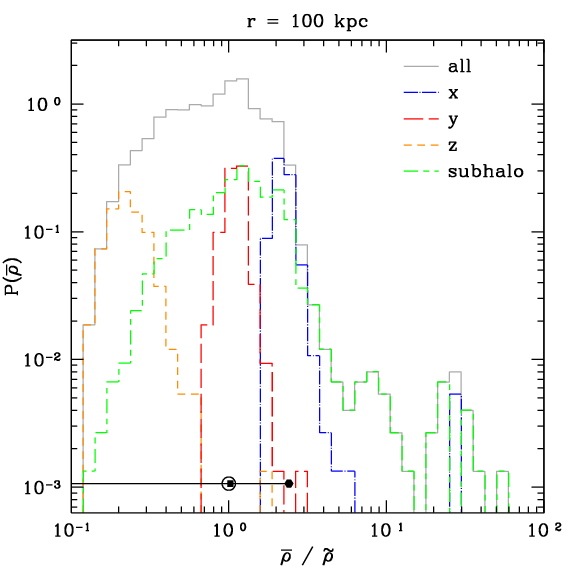}
	\includegraphics[width=0.495\textwidth,bb=0 0 574 574]{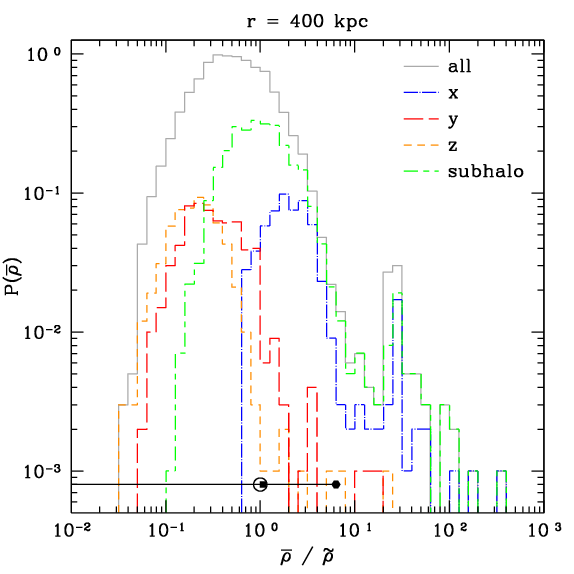}
	}
	\caption{Probability density functions of the local density $\bar{\rho}$ at different galactocentric distances $r$ normalised by the spherically averaged value $\tilde{\rho}$ (solid line/grey). Additionally, we plot the sub-samples along the different axes of the triaxial halo: $x$-axis sample (dash-dotted line/blue), $y$-axis sample (long-dashed line/red) and $z$-axis sample (short-dashed line/orange). We also plot the subhalo-affected sample (long-short-dashed line/green). The ensemble average $\left\langle \bar{\rho} \right\rangle$ over the spheres is marked with a solid black square and the standard deviation range $\left\langle \bar{\rho} \right\rangle \pm \sigma(\bar{\rho})$ is marked with a solid black hexagon. The shell value $\tilde{\rho}$ is marked with an open black circle.}
	\label{fig:rho}
\end{figure*}

In Fig. \ref{fig:rho}, we plot the probability density function of the local mean density $\bar{\rho}$ within the spheres in logarithmic scale. The local density is simply defined by the mass within a sphere divided by the volume of the sphere. In the central region, the structure of the halo looks pretty smooth in density. At 8 kpc, we find a maximum total spread of a factor $\pm$ 2 in local density with a standard deviation of $\sigma(\bar{\rho}) / \tilde{\rho} \approx 0.26$. This is due to the triaxiality of halo since the underdense spheres lie along the $z$-axis and the overdense spheres along the $x$-axis. In the outer part of the halo, however, the smooth picture does not apply. For example at 200 kpc and 400 kpc, underdense regions are found away from the short axis and the distributions of the sub-samples along the other axes are broader than in the centre of the halo. The possible range in densities at 400 kpc is close to 4 orders of magnitude with a peak probability at around $0.5~\tilde{\rho}(400~\kpc)$, i.e. the typical value for the local densities is below the spherical average value. Beyond 100 kpc, the standard deviation becomes larger than the mean value, e.g. at 400 kpc $\sigma(\bar{\rho}) / \tilde{\rho} \approx 5.3$. 

Naively, one would expect a decrease in the density variation due to shape in the outskirts of the halo, since its shape becomes rounder further out. However, this is exceeded by the steeper fall off of the density profile in the outskirts of the dark matter halo. The actual density contrast between the mean local density of the $x$-axis sample with respect to the mean local density of the $z$-axis, $C_{\rho} \equiv \left\langle \bar{\rho} \right\rangle_x / \left\langle \bar{\rho} \right\rangle_z$, increases from $C_{\rho}(8~\kpc) = 2.353$ to $C_{\rho}(400~\kpc) = 14.72$.

It is also striking that there are no high density spikes from subhaloes in the central region. This is just an artefact of our rather conservative definition of locality (i.e. the large size of the spheres). There are only very few surviving and well resolved subhaloes within 8 kpc in Via Lactea II. The subhaloes that survive close to the centre are small and compact due to tidal mass loss so that with our still relatively large definition of locality, their contribution to the mass of the spheres is small. By decreasing the size of the spheres the extremes of the distribution become more populated (see also appendix \ref{sec:comparison}). With higher resolution, one expects more subhaloes to survive which are even more compact. But one would also decrease the size of the spheres that define locality in our scheme. Hence, it is not clear what the outcome of these two competing effects would be. From simple analytical models, one expects some contribution from these small haloes \citep{2008PhRvD..77j3509K} though the details depend on their volume filling factor.

While high peaks from subhaloes are expected, underdense regions away from the short axis of the shape ellipsoid are surprising. In order to further investigate these underdensities, we repeated the measurement of local properties with spheres of 4 times smaller radii, i.e. $r_{\mathrm{sph}}/4$. In these spheres, one would expect on average approximately 21 particles in the inner region. We then simply count the small spheres that contain no particles and calculate from that the fraction of empty spheres (given in Table \ref{tab:shellsummary}). Assuming balls-in-bins statistics (see appendix \ref{sec:bib} for more details), the probability of getting empty spheres is given by $p_{\mathrm{empty}}(21) = \mathrm{e}^{-21} = 7.583 \times 10^{-10}$. Hence one would not expect to find any sphere of size $r_{\mathrm{sph}}/4$ that is completely empty. The fact that approximately 2\% of all these smaller spheres at 400 kpc are empty shows clearly that the outskirts of dark matter haloes are far from smooth in position space. Of course, to some degree this is due to the triaxial shape as actually most of the empty spheres in the outskirts are in low density regions along the $z$-axis. Taking the lower density into account, one would expect approximately 6 particles per sphere in the $z$-axis sample at 400 kpc. The probability of getting empty spheres is $p_{\mathrm{empty}}(6)  = \mathrm{e}^{-6} = 2.479 \times 10^{-3}$. The measured fraction of empty spheres in the $z$-sample is $f_{\mathrm{empty},z} = 6.656 \times 10^{-2}$ which is still clearly much higher.

We quantify this further with a statistical indicator that measures inequality: the Gini coefficient. The Gini coefficient is defined as $G \equiv 1 - 2 A$  where $A$ is the area under the Lorenz curve of the probability distribution \citep{1905PAmSA...9..209L,1912vamu.book.....G}. For the discrete probability density functions of the local mean density this results in
\begin{equation}
G(\bar{\rho}) = 1 - \frac{\sum_{i=1}^{N_{\mathrm{sample}}} p_i(\bar{\rho}) (S_i + S_{i-1})}{S_{N_{\mathrm{sample}}}}
\end{equation}
with
\begin{equation}
S_i \equiv \sum_{j=1}^{i} \bar{\rho}~p_j(\bar{\rho})~,
\end{equation}
$S_0 \equiv 0$ and
\begin{equation}
p_i(\bar{\rho}) \equiv P_i(\bar{\rho}) \Delta \log(\bar{\rho}/\tilde{\rho})~.
\end{equation}
with $\Delta \log(\bar{\rho}/\tilde{\rho})$ being the bin width in logarithmic scale. The range of the Gini coefficient is between 0 and 1. A Gini coefficient of 0 means that all the spheres have equal density whereas a Gini coefficient of 1 means that all the mass is in one sphere. As can be seen in Table \ref{tab:shellsummary}, the Gini coefficient increases with galactocentric distance showing again that the outskirts of dark matter haloes have a clumpy structure. 

This clumpy structure is due to the hierarchical build up of dark matter haloes by accretion of subhaloes. These subhaloes leave tidal streams along their orbits when they are falling into the host halo. In the outer parts of the halo we find many overlapping streams locally (see also section \ref{sec:velspace}) that originate from the many subhaloes that passed a certain region. Combined with the long local dynamical time-scale in the outskirts of haloes which inhibits effective mixing, this leads to the clumpy structure and large spread in density we see on small local scales in the outer parts of dark matter haloes.

The tidal debris can have an effect on the local dark matter annihilation \citep{2008arXiv0811.1582A}. We can quantify this by the boost factor $B \equiv \left\langle \bar{\rho}^2 \right\rangle / \left\langle \bar{\rho} \right\rangle^2$. We can only say something about local boost factors since by averaging over a spherical shell, the boost factor would be mainly driven by the triaxiality. The biggest effect is expected in the outskirts of the halo since there we find the highest degree of clumpiness. In order just to measure the influence of the debris structure and exclude the effect of the subhaloes, we only include spheres that are not subhalo-affected and get values from $B_{z,nosh}(400~\kpc) = 1.319$ to $B_{x,nosh}(400~\kpc) = 3.196$. This is in agreement with \citet{2008arXiv0811.1582A} who get of $\mathcal{O}(1)$ for the boost factor originating in the structure of the debris. If we include the subhalo-affected sample we get values in the range from $B_{y}(400~\kpc) = 6.046$ to $B_{x}(400~\kpc) = 16.15$ locally, showing showing that even in the outer halo most of the clumpiness comes from the subhalos themselves and not from their more diffuse tidal debris.

Generally, all the probability density function plots in this paper are lower limits for a possible real variation of local properties in nature since numerical effects due to numerical under-resolving (even in Via Lactea II) lead to artificial heating, smoothing, and in general loss of structure. A more detailed discussion of numerical effects follows in appendix \ref{sec:comparison}.

\subsection{Local mean velocities}\label{sec:meanvel}

\begin{figure*}
	\centering
	\ifthenelse{\boolean{useepsfigures}}{
	\includegraphics[width=0.495\textwidth]{shell8kpc_vrad}
	\includegraphics[width=0.495\textwidth]{shell25kpc_vrad}\\	
	\includegraphics[width=0.495\textwidth]{shell100kpc_vrad}
	\includegraphics[width=0.495\textwidth]{shell400kpc_vrad}
	}{
	\includegraphics[width=0.495\textwidth,bb=0 0 574 574]{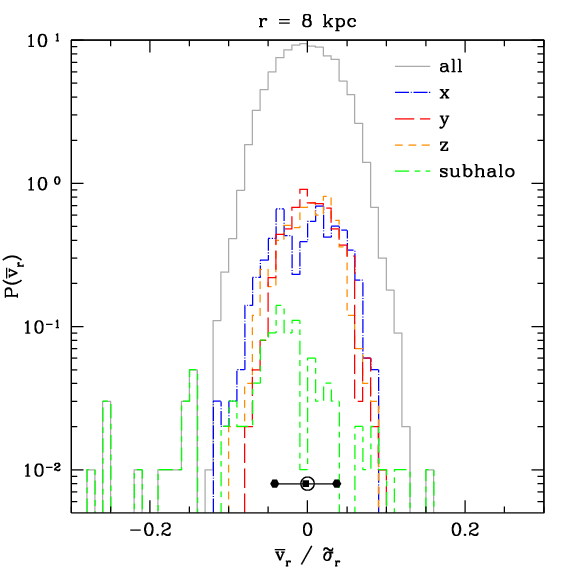}
	\includegraphics[width=0.495\textwidth,bb=0 0 574 574]{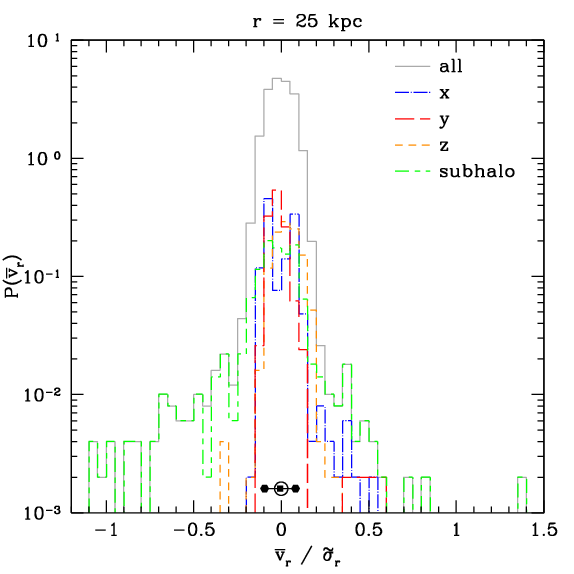}\\	
	\includegraphics[width=0.495\textwidth,bb=0 0 574 574]{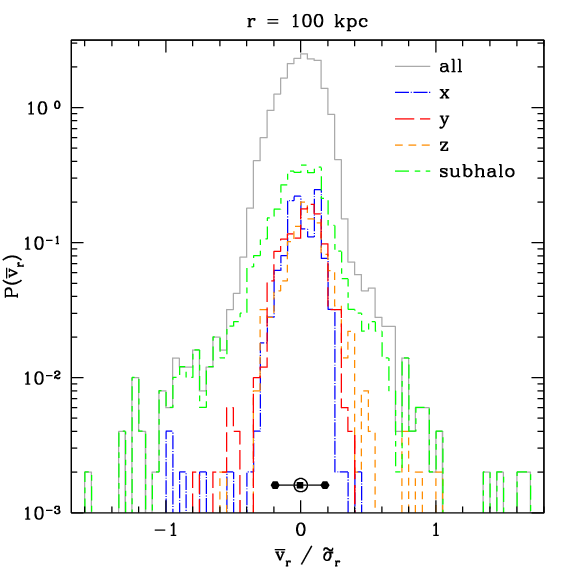}
	\includegraphics[width=0.495\textwidth,bb=0 0 574 574]{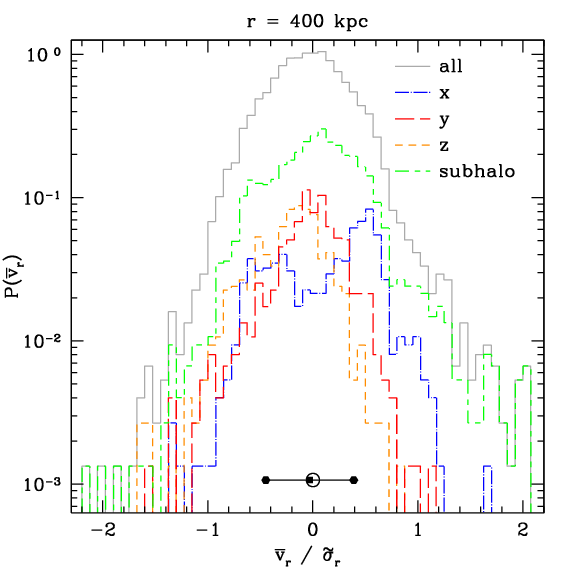}
	}
	\caption{Probability density functions of $\bar{v}_{\mathrm{r}}$ at different galactocentric distances $r$ normalised by the spherically averaged radial velocity dispersion $\tilde{\sigma}_r$. We plot the same sub-samples as in Fig. \ref{fig:rho} and use the same notation.}
	\label{fig:vrad}
\end{figure*}

\begin{figure*}
	\centering
	\ifthenelse{\boolean{useepsfigures}}{
	\includegraphics[width=0.495\textwidth]{shell8kpc_vphi}
	\includegraphics[width=0.495\textwidth]{shell25kpc_vphi}\\	
	\includegraphics[width=0.495\textwidth]{shell100kpc_vphi}
	\includegraphics[width=0.495\textwidth]{shell400kpc_vphi}
	}{
	\includegraphics[width=0.495\textwidth,bb=0 0 574 574]{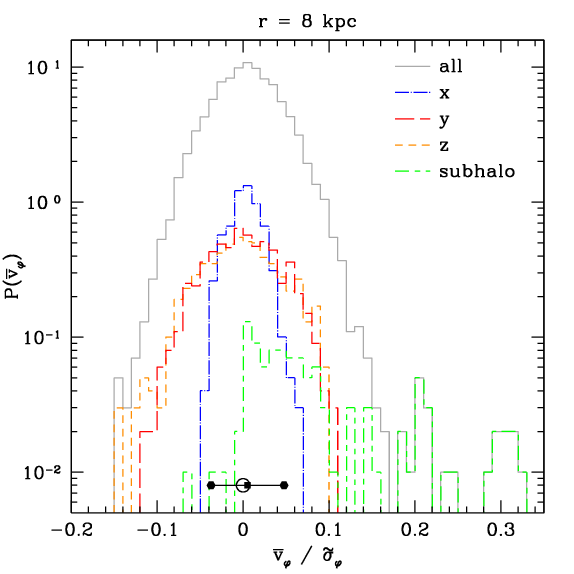}
	\includegraphics[width=0.495\textwidth,bb=0 0 574 574]{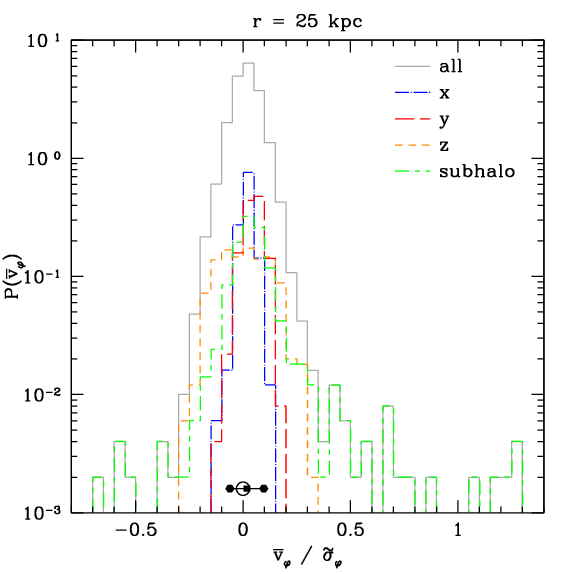}\\	
	\includegraphics[width=0.495\textwidth,bb=0 0 574 574]{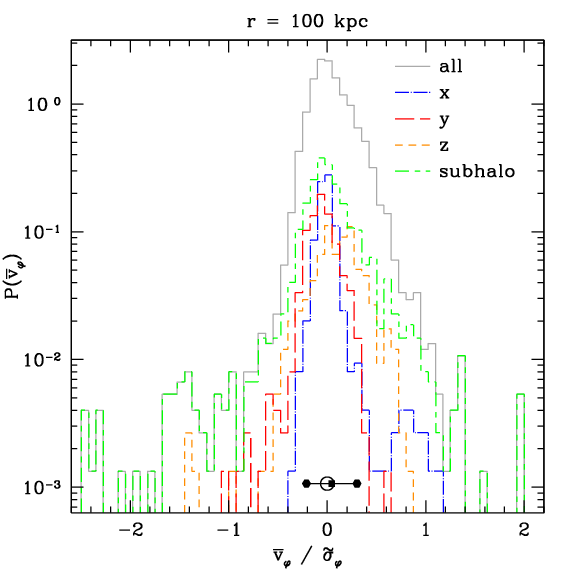}
	\includegraphics[width=0.495\textwidth,bb=0 0 574 574]{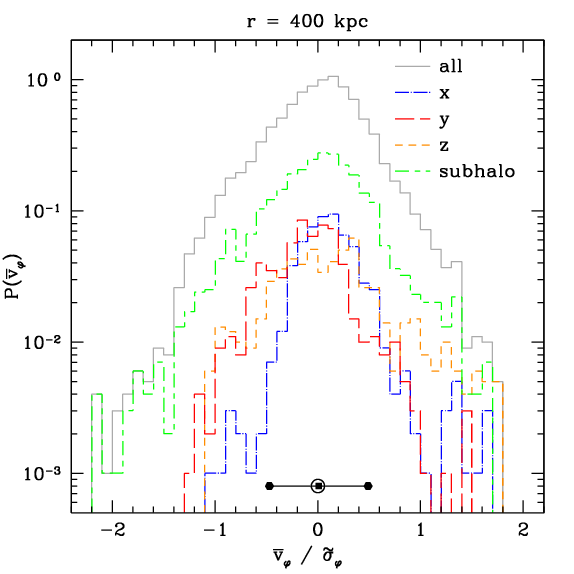}
	}
	\caption{Probability density functions of $\bar{v}_{\varphi}$ at different galactocentric distances $r$ normalised by the spherically averaged $\varphi$-velocity dispersion $\tilde{\sigma}_{\varphi}$. We plot the same sub-samples as in Fig. \ref{fig:rho} and use the same notation.}
	\label{fig:vphi}
\end{figure*}

In this section we investigate the properties of the local mean velocities, which we calculate by number weighted averages within the spheres. For this purpose, we split the velocity vector field in a radial, azimuthal ($\varphi$ defined as the angle in the $xy$-plane from the $x$-axis) and a polar ($\vartheta$ defined as the angle form the $z$-axis) component so that the unit vectors denote a right handed system with $\vec{e}_r \wedge \vec{e}_\varphi = \vec{e}_\vartheta$. In Fig. \ref{fig:vrad} and \ref{fig:vphi}, we plot the probability density functions for the radial respectively $\varphi$-component of the velocity field in linear scale; omitting the plot for the mean $\vartheta$-velocity as it does not show qualitatively different features than the mean $\varphi$-velocity. 

Often it is assumed that bulk velocities in dark matter haloes are zero. Again, this is approximately true in the central part. The standard deviation of the local mean radial velocity is about 4\% of the shell averaged radial velocity dispersion. In the outer parts, not even the spherically averaged values are zero (see also Table \ref{tab:shellsummary}) and the local bulk velocities reach up to twice the values of the local velocity dispersion. At 400 kpc, the dispersion of the mean radial velocity exceeds 40\% of the shell averaged radial velocity dispersion. 

At different radii one can see two populations of spheres: one that is infalling and another one that is moving outwards. This is best seen along the $x$-axis but often the two populations are too much smeared out. Such a distribution naturally arises from a cosmological infall pattern and has been seen before in numerical simulations \citep{2008ApJ...680L..25D}.

A similar picture emerges from the mean $\varphi$- and $\vartheta$-velocities with a comparable spread of the probability density distributions at a given distance as for the mean radial velocities. We find that the sub-population of spheres along the long axis have a narrower and more centrally peaked distribution in mean $\varphi$- and $\vartheta$-velocities than the sub-populations along the short and intermediate axes indicating some degree of coherent tangential flow around the short and intermediate axes.

The distribution of the subhalo-affected sample of spheres is always much broader than the total sample, and it contributes most to the extremes (both high and low) of the overall distribution. This indicates that the subhalo population as a whole is kinematically hotter than the background, in agreement with earlier studies by \cite{2004MNRAS.352..535D}. Indeed we find subhaloes in the inner halo that have a speed of around $500~\km~\s^{-1} \approx 3.5~\sigma_{\mathrm{r}}(8~\kpc)$. 

In the inner region, we find that the subhalo-affected sample is skewed towards negative radial velocities (i.e. infalling spheres) and positive $\varphi$-velocities. The skew in radial velocity is a signature of the tidal disruption process: we see these subhaloes on their last infall before they either are completely disrupted or they loose so much mass that they do not contribute much to the sphere averaged properties any more on their way out since they are now much more compact and less massive.

All this conclusively shows that the traditional notion of a dark matter halo with no local bulk velocities is an inaccurate description for structures that form in cosmological simulations, especially in the outer parts of haloes.

\subsection{Local velocity dispersions}

\begin{figure*}
	\centering
	\ifthenelse{\boolean{useepsfigures}}{
	\includegraphics[width=0.495\textwidth]{shell8kpc_sigtot}
	\includegraphics[width=0.495\textwidth]{shell25kpc_sigtot}\\
	\includegraphics[width=0.495\textwidth]{shell100kpc_sigtot}
	\includegraphics[width=0.495\textwidth]{shell400kpc_sigtot}
	}{
	\includegraphics[width=0.495\textwidth,bb=0 0 574 574]{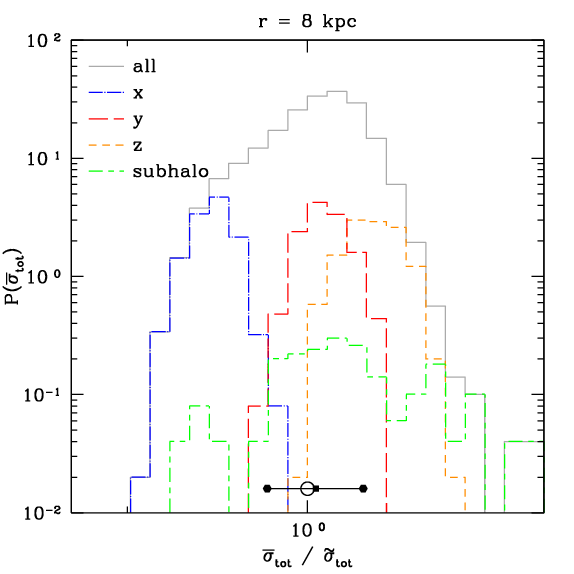}
	\includegraphics[width=0.495\textwidth,bb=0 0 574 574]{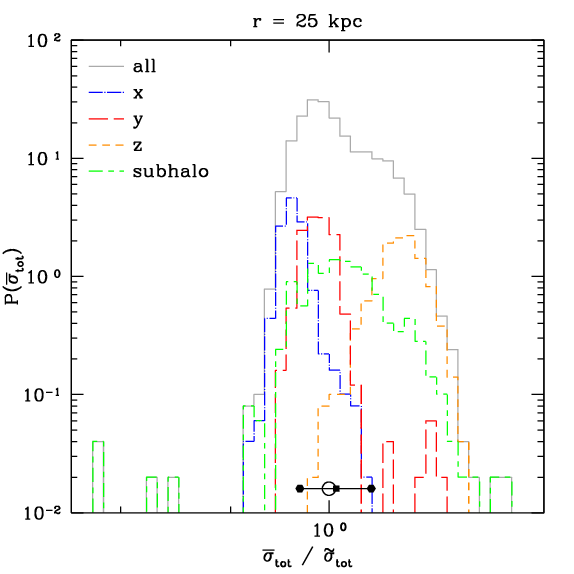}\\
	\includegraphics[width=0.495\textwidth,bb=0 0 574 574]{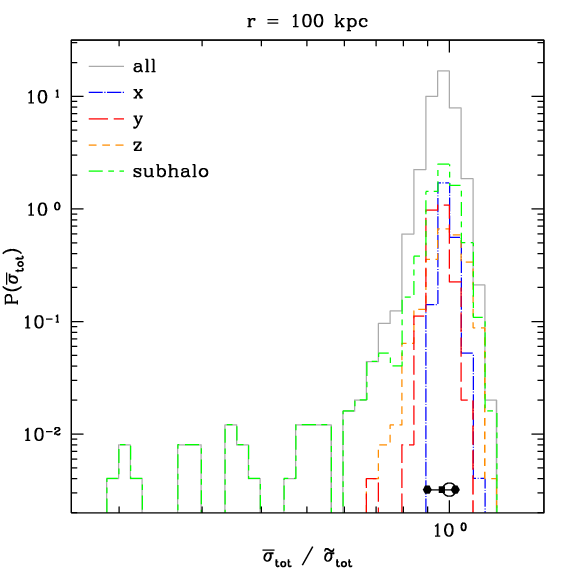}
	\includegraphics[width=0.495\textwidth,bb=0 0 574 574]{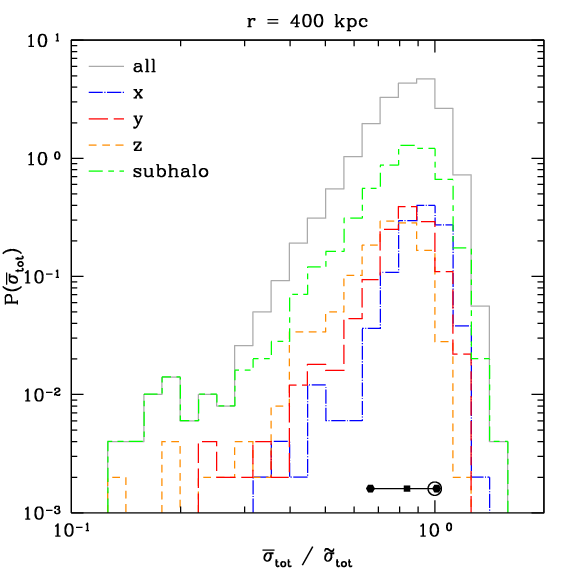}
	}
	\caption{Probability density functions of the total velocity dispersion $\bar{\sigma}_{\mathrm{tot}}$ at different galactocentric distances $r$ normalised to the spherically averaged value $\tilde{\sigma}_{\mathrm{tot}}$. We plot the same sub-samples as in Fig. \ref{fig:rho} and use the same notation.}
	\label{fig:sigtot}
\end{figure*}

\begin{figure*}
	\centering
	\ifthenelse{\boolean{useepsfigures}}{
	\includegraphics[width=0.495\textwidth]{shell8kpc_sigrad}
	\includegraphics[width=0.495\textwidth]{shell25kpc_sigrad}\\
	\includegraphics[width=0.495\textwidth]{shell100kpc_sigrad}
	\includegraphics[width=0.495\textwidth]{shell400kpc_sigrad}
	}{
	\includegraphics[width=0.495\textwidth,bb=0 0 574 574]{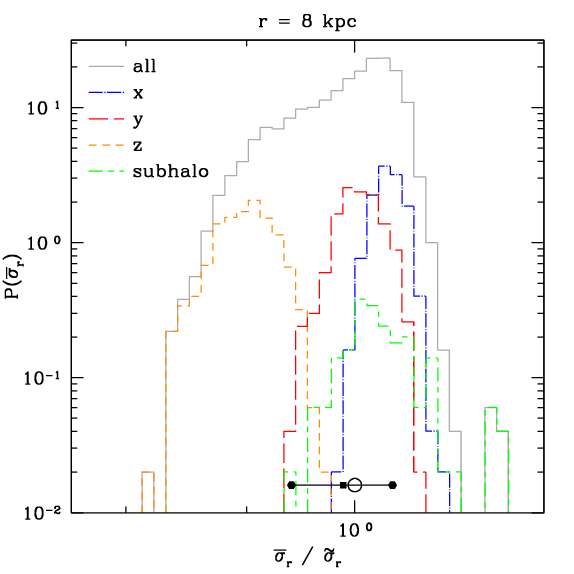}
	\includegraphics[width=0.495\textwidth,bb=0 0 574 574]{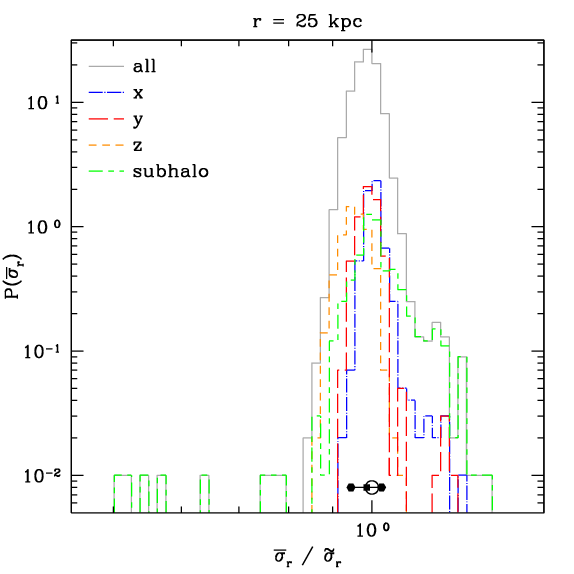}\\
	\includegraphics[width=0.495\textwidth,bb=0 0 574 574]{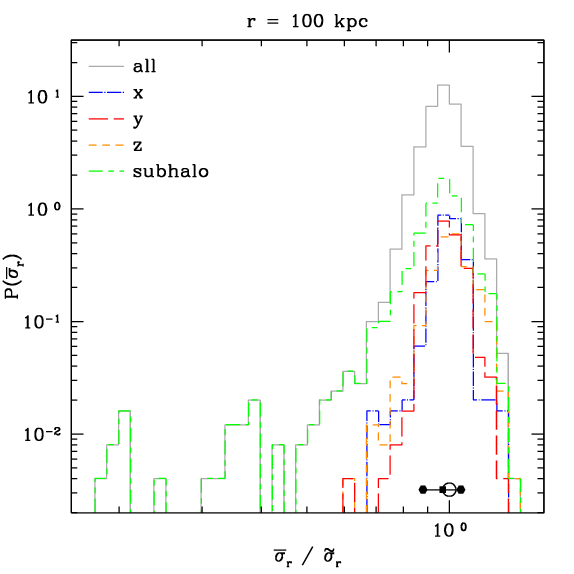}
	\includegraphics[width=0.495\textwidth,bb=0 0 574 574]{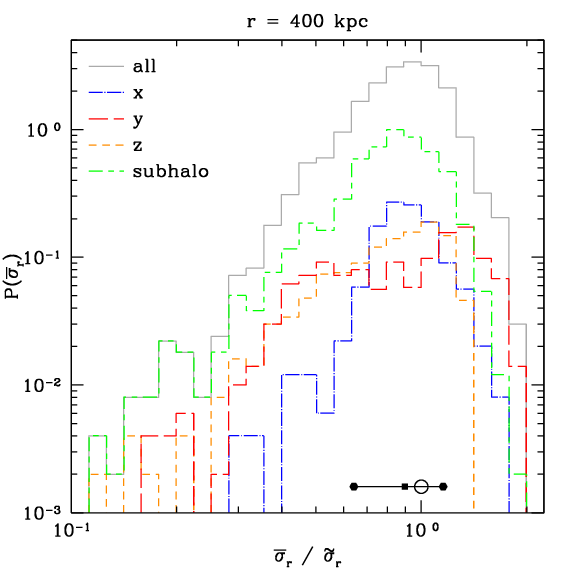}
	}
	\caption{Probability density functions of radial velocity dispersion $\bar{\sigma}_r$ at different galactocentric distances $r$ normalised to the spherically averaged value $\tilde{\sigma}_r$. We plot the same sub-samples as in Fig. \ref{fig:rho} and use the same notation.}
	\label{fig:sigrad}
\end{figure*}

\begin{figure*}
	\centering
	\ifthenelse{\boolean{useepsfigures}}{
	\includegraphics[width=0.495\textwidth]{shell8kpc_sigphi}
	\includegraphics[width=0.495\textwidth]{shell25kpc_sigphi}\\
	\includegraphics[width=0.495\textwidth]{shell100kpc_sigphi}
	\includegraphics[width=0.495\textwidth]{shell400kpc_sigphi}
	}{
	\includegraphics[width=0.495\textwidth,bb=0 0 574 574]{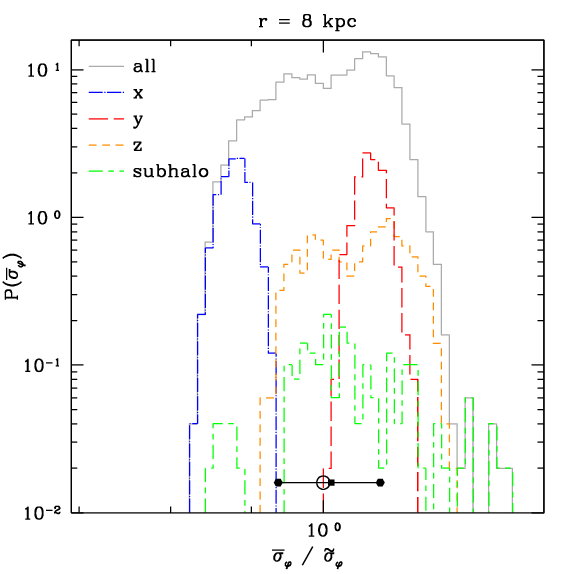}
	\includegraphics[width=0.495\textwidth,bb=0 0 574 574]{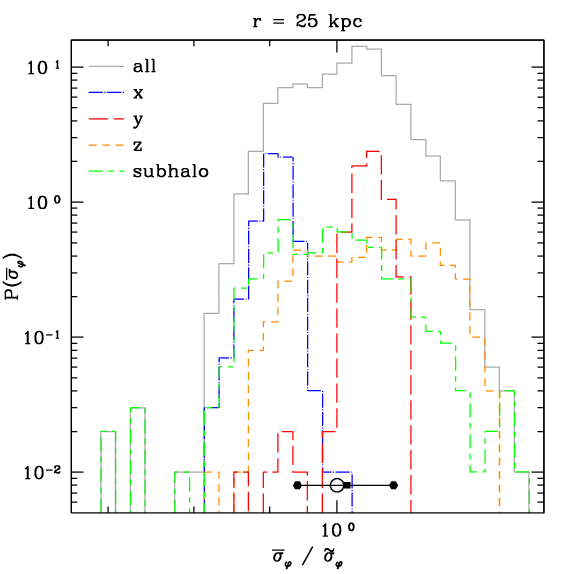}\\
	\includegraphics[width=0.495\textwidth,bb=0 0 574 574]{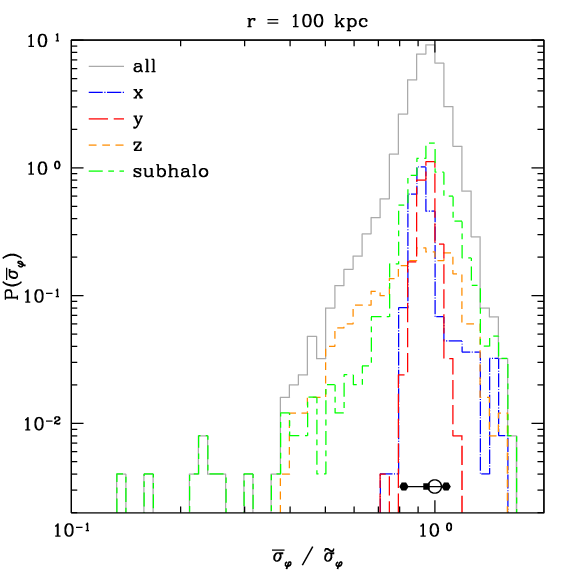}
	\includegraphics[width=0.495\textwidth,bb=0 0 574 574]{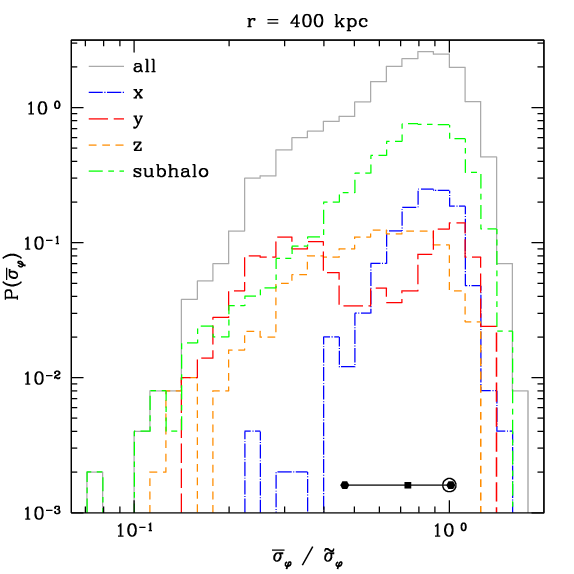}
	}
	\caption{Probability density functions of $\varphi$-velocity dispersion $\bar{\sigma}_\varphi$ at different galactocentric distances $r$ normalised to the spherically averaged value $\tilde{\sigma}_\varphi$. We plot the same sub-samples as in Fig. \ref{fig:rho} and use the same notation.}
	\label{fig:sigphi}
\end{figure*}

In Fig. \ref{fig:sigtot}, \ref{fig:sigrad} and \ref{fig:sigphi} we plot the probability density functions of the local total, radial and $\varphi$-velocity dispersion. We omit the figure for the $\vartheta$-velocity dispersion since it shows a similar behaviour as the $\varphi$-velocity dispersion. The local velocity dispersion is given by $\bar{\sigma}_k^2 = \overline{v_k^2} - \bar{v}_k^2$ for $k = r, \varphi ~\mathrm{or}~\vartheta$ and $\bar{\sigma}_{\mathrm{tot}}^2 = \bar{\sigma}_a^2 + \bar{\sigma}_b^2 + \bar{\sigma}_c^2$, where $\sigma_a$, $\sigma_b$ respectively $\sigma_c$ are the values of the dispersions in the eigencoordinate system of the local dispersion ellipsoid (see section \ref{sec:veldispellipsoid} for more details). 

We see that the local velocity dispersion has strong underlying variations. Whereas the central part only has variations on the few per cent level with respect to the spherically averaged value (e.g. at 8 kpc we have $\sigma(\bar{\sigma}_r) / \tilde{\sigma}_r \approx 0.05$), one can find regions in the outer part of the halo that deviate by approximately an order of magnitude in velocity dispersion and at 400 kpc we have for example $\sigma(\bar{\sigma}_r) / \tilde{\sigma}_r \approx 0.25$. 

The total velocity dispersion shows the most compact probability density function when measured with respect to the spherically averaged value, i.e. $\sigma(\bar{\sigma}_{\mathrm{tot}}) / \tilde{\sigma}_{\mathrm{tot}}$ is smaller than $\sigma(\bar{\sigma}_k) / \tilde{\sigma}_k$ for $k = r, \varphi ~\mathrm{or}~\vartheta$ and the distributions peak generally around the spherically averaged value. 

Interestingly, we find that regions in the centre along the long axis are colder (i.e. $\bar{\sigma}_{\mathrm{tot}}$ is smaller) than spheres along the intermediate or short axes. This trend is not observed in the outskirts of the halo. We also find that along the long axis, the radial velocity dispersions are higher and the tangential velocity dispersions are low in the inner region of the halo. This is due to the prolate shape and orientation of the local velocity dispersion ellipsoid and is discussed in more detail in section \ref{sec:veldispellipsoid}. In general, we find that the distributions of velocity dispersions for the two tangential velocity components are much broader along the short axis than along the long and intermediate axes. The subhalo-affected sample shows a less peaked distribution in velocity dispersions than the total sample. In the regions with enough subhaloes, the distribution is skewed towards lower dispersions for all velocity components. This is of course due to the lower internal velocity dispersion in subhaloes.

It remains to be seen if these empirical trends of velocity dispersion with orientation are universal for dark matter haloes or if they depend on the detailed hierarchical build-up history of every individual halo. Nevertheless, it is clear that the degree of chaotic motion can strongly vary locally and deviate by substantial factors from the spherically averaged value.

\subsection{Local anisotropy parameter}\label{sec:ani}

\begin{figure*}
	\centering
	\ifthenelse{\boolean{useepsfigures}}{
	\includegraphics[width=0.495\textwidth]{shell8kpc_ani}
	\includegraphics[width=0.495\textwidth]{shell25kpc_ani}\\
	\includegraphics[width=0.495\textwidth]{shell100kpc_ani}
	\includegraphics[width=0.495\textwidth]{shell400kpc_ani}
	}{
	\includegraphics[width=0.495\textwidth,bb=0 0 574 574]{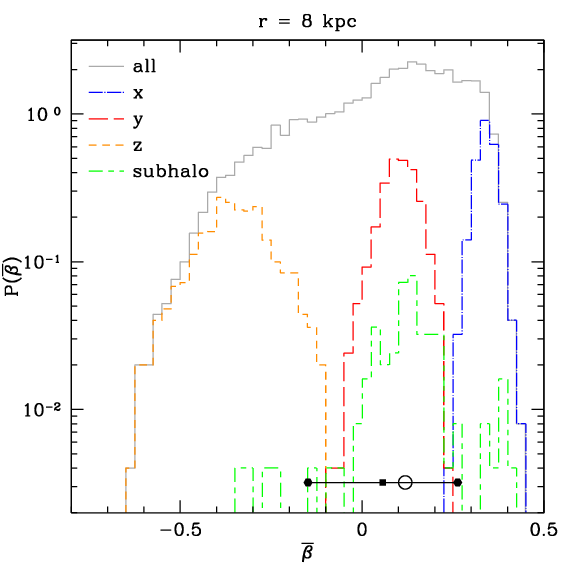}
	\includegraphics[width=0.495\textwidth,bb=0 0 574 574]{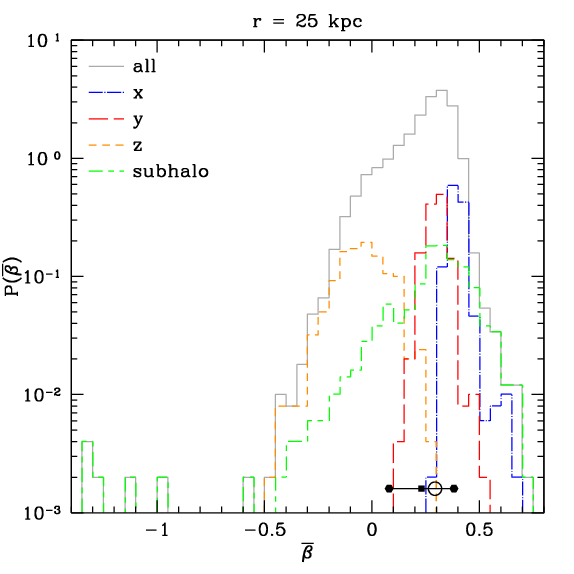}\\
	\includegraphics[width=0.495\textwidth,bb=0 0 574 574]{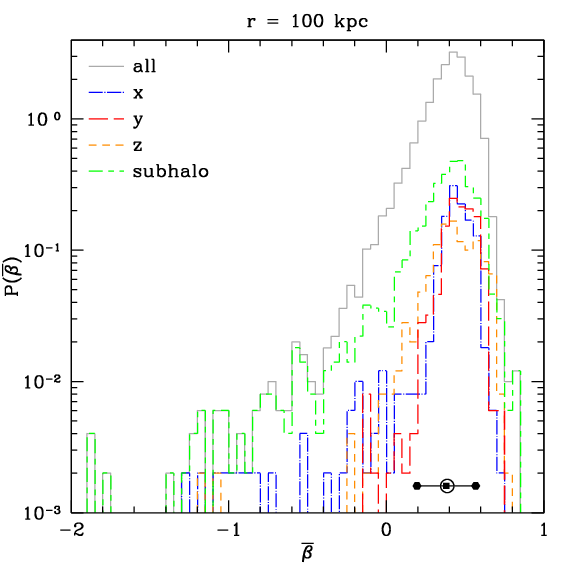}
	\includegraphics[width=0.495\textwidth,bb=0 0 574 574]{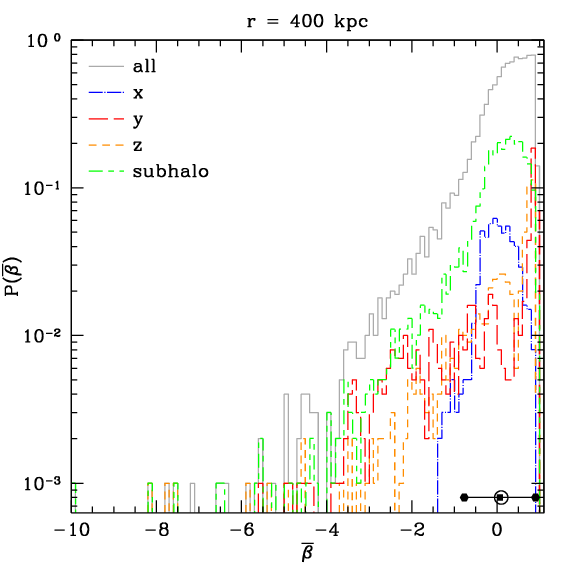}
	}
	\caption{Probability density functions of the local anisotropy parameter $\bar{\beta}$ at different galactocentric distances $r$. We plot the same sub-samples as in Fig. \ref{fig:rho} and use the same notation.}
	\label{fig:ani}
\end{figure*}

In Fig. \ref{fig:ani}, we plot the probability density function of the local anisotropy parameter $\bar{\beta}$ within the spheres. The local anisotropy parameter is given by
\begin{equation}
\bar{\beta} = 1 - \frac{1}{2} \frac{\bar{\sigma}_t^2}{\bar{\sigma}_r^2}
\end{equation}
where $\bar{\sigma}_t^2 = \bar{\sigma}_\varphi^2 + \bar{\sigma}_\vartheta^2$ is the tangential velocity dispersion squared and where we neglect the correlation term. 

This form of the anisotropy parameter is often used but has some problems since the following assumptions which are hidden in this expression are in general not fulfilled: i) $\overline{v_\varphi^2} = \overline{v_\vartheta^2}$, ii) $\bar{v}_r = \bar{v}_\varphi = \bar{v}_\vartheta = 0$ and iii) $\mathrm{cov}(v_\varphi,v_\vartheta) = 0$. These three conditions are only approximately fulfilled in the central part of a halo and certainly not in the outskirts. One should therefore see the expression for $\bar{\beta}$ as a definition for local anisotropy so that a comparison with previous work is possible.

We find that the radial anisotropy of the velocity dispersion tensor in our halo increases with radius (see also Table \ref{tab:shellsummary}), in agreement with previous studies \citep[see e.g.][]{1996MNRAS.281..716C,2000ApJ...539..561C,2001ApJ...557..533F,2004MNRAS.352..535D,2006NewA...11..333H}. In the central part we find that regions along the long axis are preferentially on radial orbits, whereas regions along the short axis are tendencially on tangential orbits. This effect disappears at larger radii. It can be understood as a direct consequence of the variation of the orientation of the local velocity dispersion ellipsoid discussed in the next section \ref{sec:veldispellipsoid}.

The probability densities peak in general close to the spherical average value except at 400 kpc where it is peaked towards highly radial anisotropies. This is especially true for the the sub-profiles of the $y$- and $z$-axis sample whereas the $x$-axis sample peaks around $\bar{\beta} = 0$. With increasing distance also the spread in local anisotropy becomes larger and can vary from close to perfectly radial to close to perfectly tangential. By inspecting the subhalo-affected sample, we find that this sample is skewed towards tangential values of $\bar{\beta}$ at galactocentric distances with enough subhaloes.

\subsection{Local velocity dispersion ellipsoid}\label{sec:veldispellipsoid}

\begin{figure*}
	\centering
	\ifthenelse{\boolean{useepsfigures}}{
	\includegraphics[width=0.495\textwidth]{shell8kpc_alphaa}
	\includegraphics[width=0.495\textwidth]{shell25kpc_alphaa}\\
	\includegraphics[width=0.495\textwidth]{shell100kpc_alphaa}
	\includegraphics[width=0.495\textwidth]{shell400kpc_alphaa}	
	}{
	\includegraphics[width=0.495\textwidth,bb=0 0 574 574]{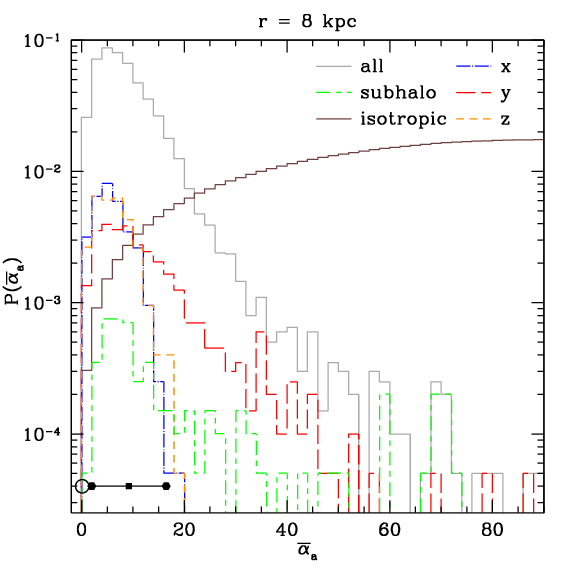}
	\includegraphics[width=0.495\textwidth,bb=0 0 574 574]{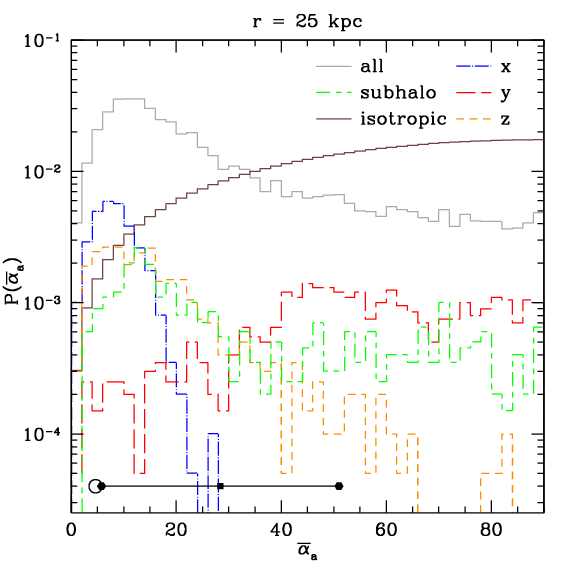}\\
	\includegraphics[width=0.495\textwidth,bb=0 0 574 574]{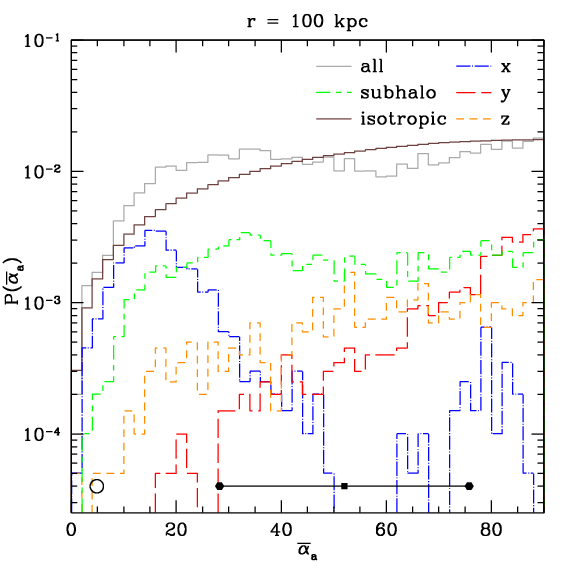}
	\includegraphics[width=0.495\textwidth,bb=0 0 574 574]{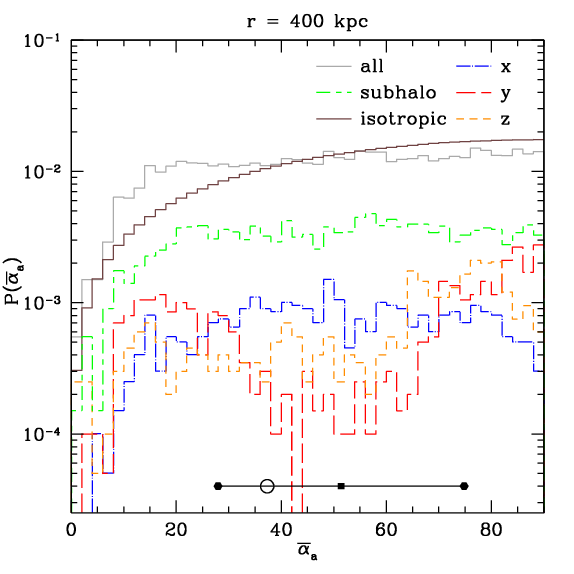}	
	}
	\caption{Probability density functions of the angle $\bar{\alpha}_a$ at different galactocentric distances $r$. We plot the same sub-samples as in Fig. \ref{fig:rho} and use the same notation.}
	\label{fig:alpha_a}
\end{figure*}

\begin{figure*}
	\centering
	\ifthenelse{\boolean{useepsfigures}}{
	\includegraphics[width=0.495\textwidth]{shell8kpc_alphab}
	\includegraphics[width=0.495\textwidth]{shell25kpc_alphab}\\
	\includegraphics[width=0.495\textwidth]{shell100kpc_alphab}
	\includegraphics[width=0.495\textwidth]{shell400kpc_alphab}	
	}{
	\includegraphics[width=0.495\textwidth,bb=0 0 574 574]{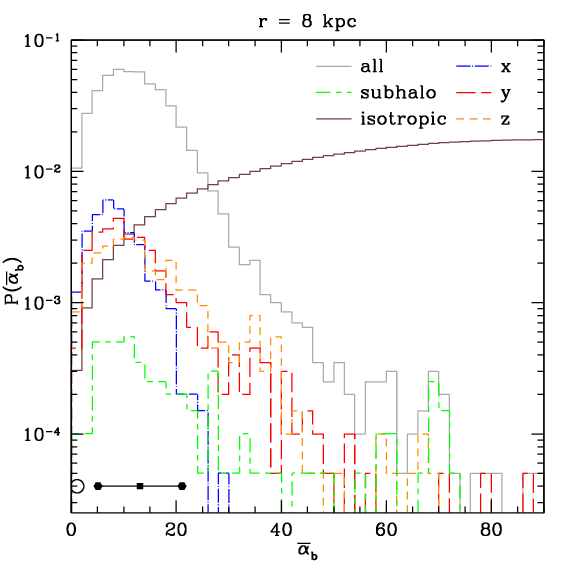}
	\includegraphics[width=0.495\textwidth,bb=0 0 574 574]{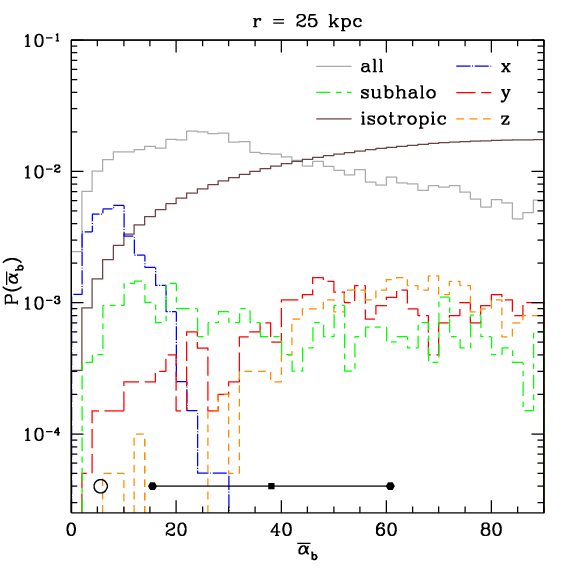}\\
	\includegraphics[width=0.495\textwidth,bb=0 0 574 574]{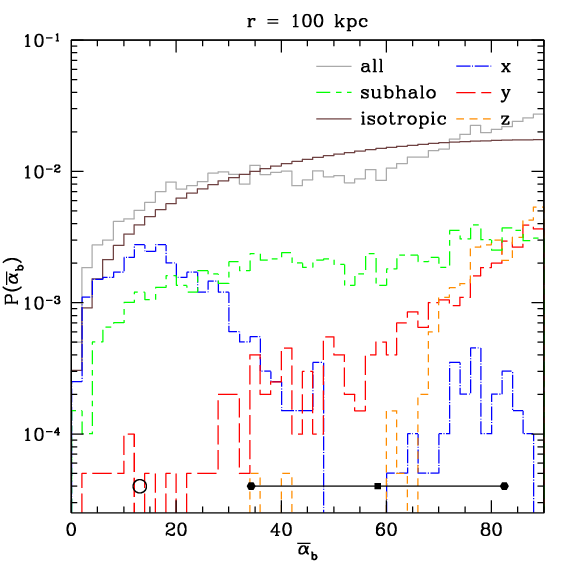}
	\includegraphics[width=0.495\textwidth,bb=0 0 574 574]{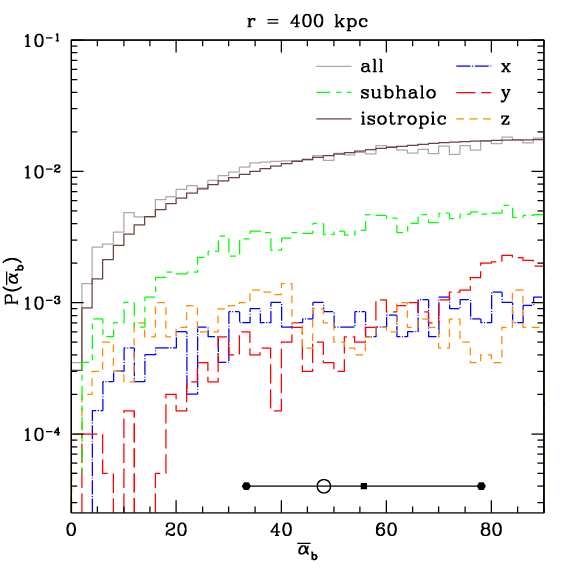}	
	}
	\caption{Probability density functions of the angle $\bar{\alpha}_b$ at different galactocentric distances $r$. We plot the same sub-samples as in Fig. \ref{fig:rho} and use the same notation.}
	\label{fig:alpha_b}
\end{figure*}

\begin{figure*}
	\centering
	\ifthenelse{\boolean{useepsfigures}}{
	\includegraphics[width=0.495\textwidth]{shell8kpc_T}
	\includegraphics[width=0.495\textwidth]{shell25kpc_T}\\
	\includegraphics[width=0.495\textwidth]{shell100kpc_T}
	\includegraphics[width=0.495\textwidth]{shell400kpc_T}
	}{
	\includegraphics[width=0.495\textwidth,bb=0 0 574 574]{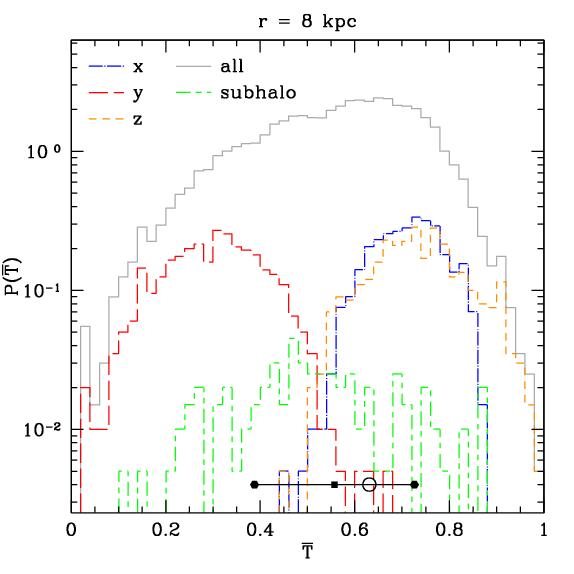}
	\includegraphics[width=0.495\textwidth,bb=0 0 574 574]{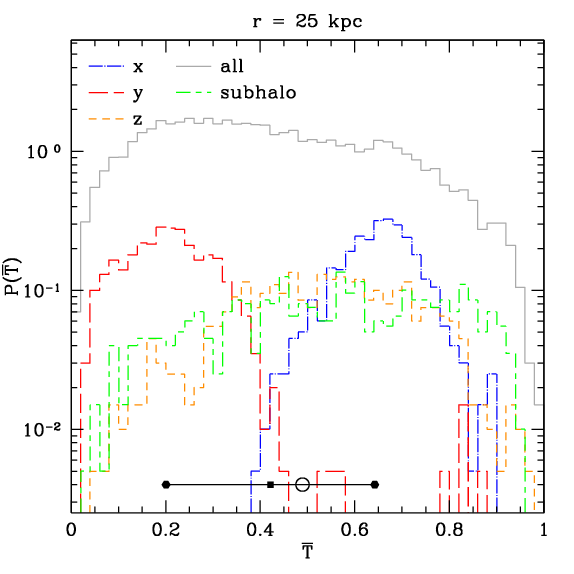}\\
	\includegraphics[width=0.495\textwidth,bb=0 0 574 574]{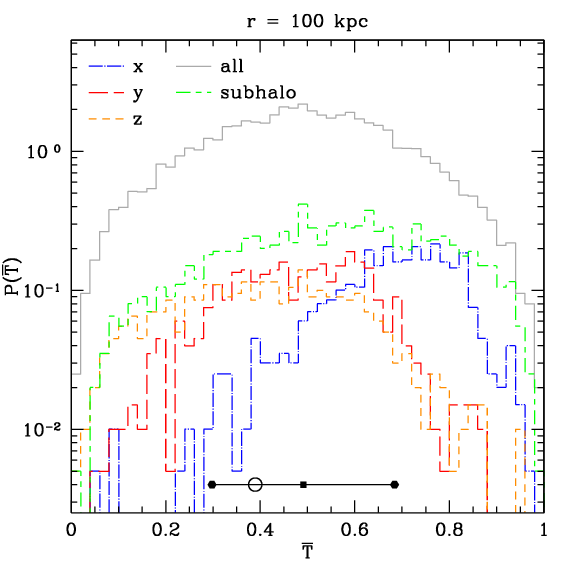}
	\includegraphics[width=0.495\textwidth,bb=0 0 574 574]{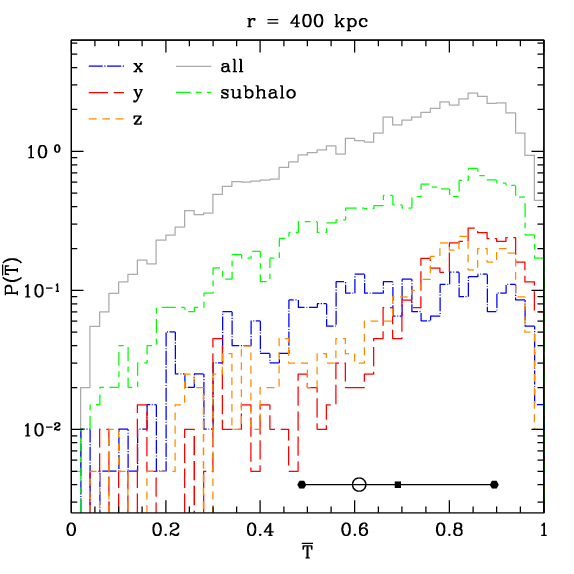}
	}
	\caption{Probability density functions of the triaxiality parameter $\overline{T}$ at different galactocentric distances $r$. We plot the same sub-samples as in Fig. \ref{fig:rho} and use the same notation.}
	\label{fig:T}
\end{figure*}

In each sphere we also calculate the local velocity covariance tensor, also know as the velocity dispersion tensor, $\mathbf{\Sigma}$ given by
\begin{equation}
\Sigma_{ij} \equiv \overline{(v_i - \bar{v}_i)(v_j - \bar{v}_j)}~.
\end{equation}
By diagonalising $\mathbf{\Sigma}$, the velocity dispersions in the eigencoordinate system are given by the square roots of the eigenvalues of the velocity covariance tensor. These velocity dispersions in the eigencoordinate system define the local velocity dispersion ellipsoid and are sorted by $\bar{\sigma}_a \geq \bar{\sigma}_b \geq \bar{\sigma}_c$. We name the appropriate eigenvectors accordingly, e.g. $\vec{e}_{\sigma_a}$ etc.

First, we check the orientation of the local velocity dispersion ellipsoid. For this purpose we calculate the angles between the eigenvectors of the velocity covariance tensor and the appropriate shape axis given by
\begin{eqnarray}
\bar{\alpha}_a & = & \arccos(\vec{e}_x \cdot \vec{e}_{\sigma_a})\\
\bar{\alpha}_b & = & \arccos(\vec{e}_y \cdot \vec{e}_{\sigma_b})\\
\bar{\alpha}_c & = & \arccos(\vec{e}_z \cdot \vec{e}_{\sigma_c})
\end{eqnarray}
In Fig. \ref{fig:alpha_a} respectively \ref{fig:alpha_b} we plot the probability density function of $\bar{\alpha}_a$ and $\bar{\alpha}_b$. We do not show the figure for $\bar{\alpha}_c$ since it shows a qualitatively similar behaviour as for $\bar{\alpha}_b$. We also plot the curve, which is proportional to $\sin(\bar{\alpha}_k)$, that corresponds to an isotropic distribution.

We observe that in the inner regions the local velocity dispersion ellipsoid is close to perfectly aligned with the shape ellipsoid since all angle probability density functions peak sharply at small angles. This effect is most prominent for the major axis sub-sample, which has a very tight and sharp distribution around small angles $\bar{\alpha}_k$ for all $k$, and less so for the intermediate and short axes, which in some cases show a broader spread towards larger angles. The general alignment of the local velocity dispersion ellipsoid with the shape of the dark matter halo nicely explains the observed behaviour of the anisotropy parameter $\bar{\beta}$, for which we mainly found radial orbits along the $x$-axis, isotropic orbits along the intermediated $y$-axis and tangential orbits along the $z$-axis. The further out we go, the more isotropic the different angle distributions become. However, at nearly all distances from the Galaxy centre we find that smaller angles are slightly more probable than in a perfectly isotropic distribution. 

Interesting is the behaviour of the different sub-samples further out. The alignment with the shape seems to be best along the $x$-axis. Only in the outskirts at 400 kpc, this sub-sample distribution becomes more and more isotropic for all angles. The sub-samples along the other two axes show even in the inner part a broader distribution that does not follow the overall distribution. For example, the $z$-axis sample shows a distribution peaked around large angles for $\bar{\alpha}_b$ and $\bar{\alpha}_c$ for intermediate distances around 50 kpc and 100 kpc whereas the $\bar{\alpha}_a$ distribution is mildly peaked towards small angles at 50 kpc (not shown in figure). In other words: only the long axis seems to be slightly aligned with the long axis of the shape ellipsoid whereas the the intermediate and short axes are perpendicular to the corresponding shape axes. In general, also the subhalo sample shows a rather isotropic distribution although it also tends to be aligned with the shape ellipsoid in the inner part of the halo. 

{\it Globally}, we expect such an alignment from the tensor virial theorem \citep{1987gady.book.....B} and such a correlation has previously been found in cosmological $N$-body simulations, e.g. in \cite{2006MNRAS.367.1781A}. If we calculate the velocity dispersion ellipsoid for all the particles in the shell, we also find a very tight alignment of this shell dispersion ellipsoid with the shape ellipsoid (see e.g. Fig. \ref{fig:alpha_a} respectively \ref{fig:alpha_b}). Only in the outer parts at 400 kpc we get a deviation from that behaviour. But in principle, we do not expect that the tensor virial theorem holds locally. It seems that the alignment of the local velocity dispersion ellipsoid with the shape ellipsoid only holds in relaxed and well mixed regions - the central region of the Via Lactea II halo. Hence, this local alignment might just be a numerical artefact since the central region is too relaxed due to numerical under-resolving if compared to the true degree of relaxation in reality. But an alignment of the local dispersion ellipsoid with the shape is also found in observations. For example, galaxies from the SAURON survey that were dynamically modelled with the Schwarzschild technique also show such a correlation \citep{2007MNRAS.379..418C}.

Furthermore, we calculate the shape of the local velocity dispersion ellipsoids which we measure with the triaxiality parameter \citep{1991ApJ...383..112F} defined by
\begin{equation}
\overline{T} \equiv \frac{\bar{\sigma}_a^2 - \bar{\sigma}_b^2}{\bar{\sigma}_a^2 - \bar{\sigma}_c^2}~.
\end{equation}
Ellipsoids are called oblate if $0 \leq \overline{T} \leq 1/3$, triaxial if $1/3 < \overline{T} < 2/3$ and prolate if $2/3 \leq \overline{T} \leq 1$. In Fig. \ref{fig:T} we plot the probability density function of the triaxiality parameter in linear scale.

The total probability density function shifts from a peak in the oblate region (e.g. at 25 kpc) via a peak in the triaxial region (e.g. at 100 kpc) to a more prolate shape at 400 kpc. Only the innermost region at 8 kpc does not follow this overall trend. Interesting is again the distribution of the different sub-samples. The shape of the velocity dispersion ellipsoids along the $x$-axis generally peak in the prolate region although the distribution becomes broader towards oblate shapes further out. The $y$-axis sample has in the inner region a rather oblate shape which drifts via triaxial (e.g. at 100 kpc) to a prolate distribution in the outskirts of the halo. Completely different is the behaviour of the $z$-axis sample: at 8 kpc it's distribution is peaked in the prolate region then drifts via the triaxial region at 25 kpc to a oblate distribution at 50 kpc (not shown in the figure). Then it swings back via a rather triaxial shape to a prolate shape again at 400 kpc. The shape of the shell averaged velocity dispersion tensor does not fully follow the trend of the overall distribution and is in general not close to the peak of the total probability density function.

At the moment it is not clear what the underlying cause for these shape and orientation variations with galactocentric distance is and if these trends are universal. Further numerical investigations are necessary.

\subsection{Local velocity space}\label{sec:velspace}

\begin{figure*}
	\centering
	\includegraphics[width=0.315\textwidth,bb=0 0 1024 1024]{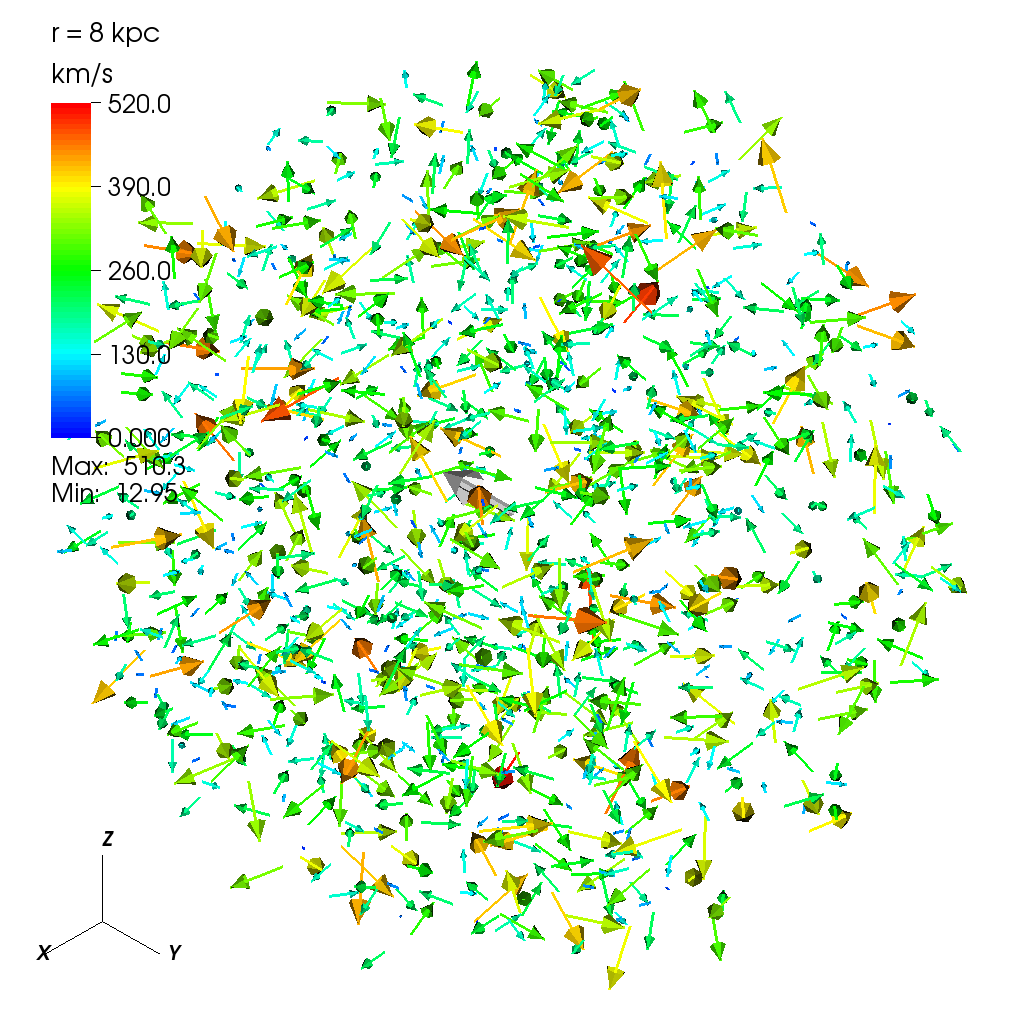}
	\includegraphics[width=0.315\textwidth,bb=0 0 1024 1024]{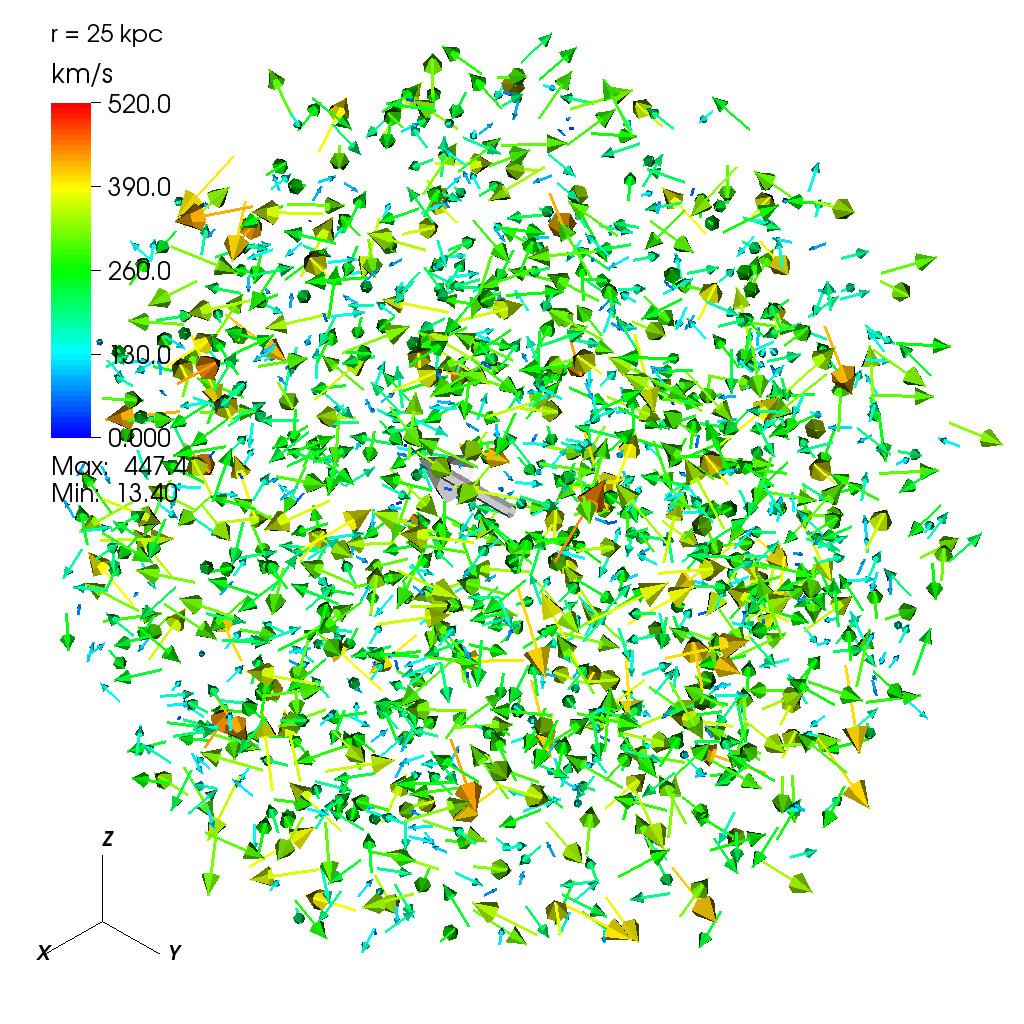}
	\includegraphics[width=0.315\textwidth,bb=0 0 1024 1024]{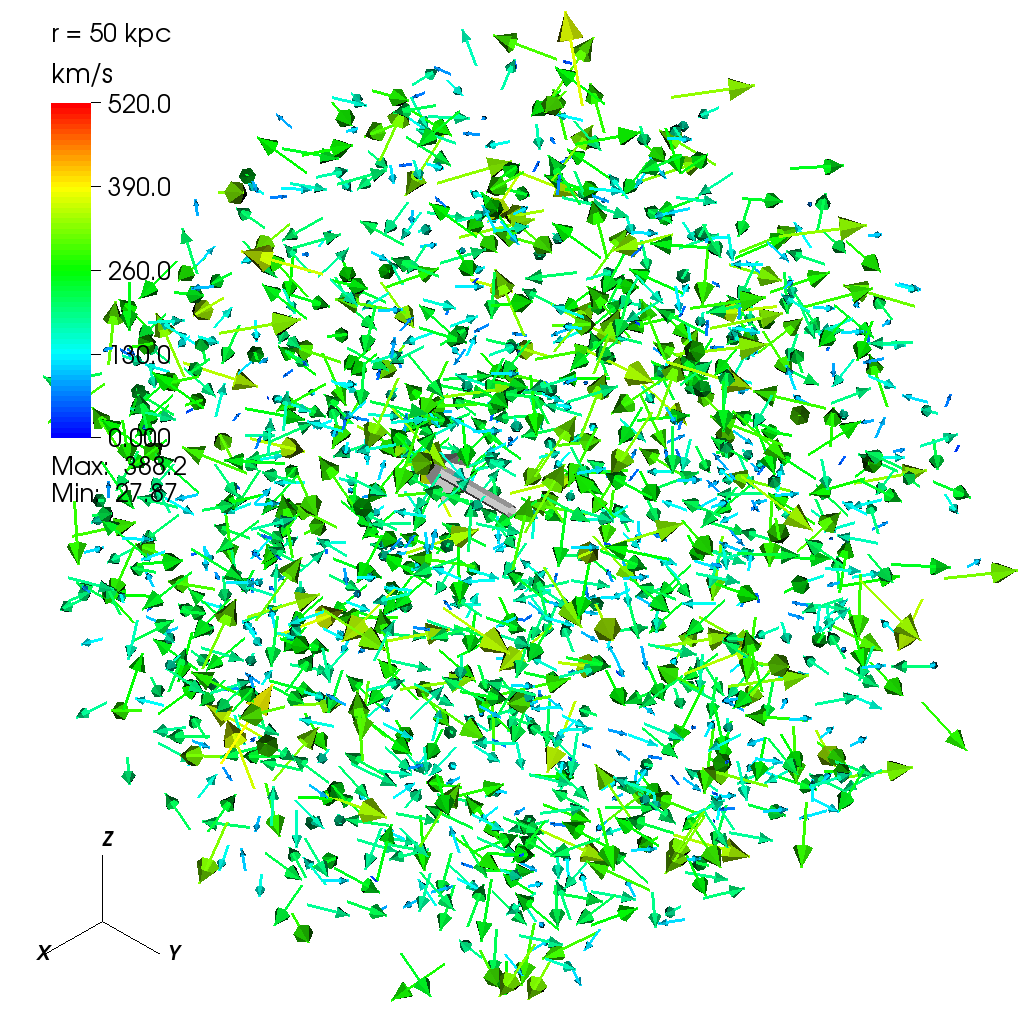}\\
	\includegraphics[width=0.315\textwidth,bb=0 0 1024 1024]{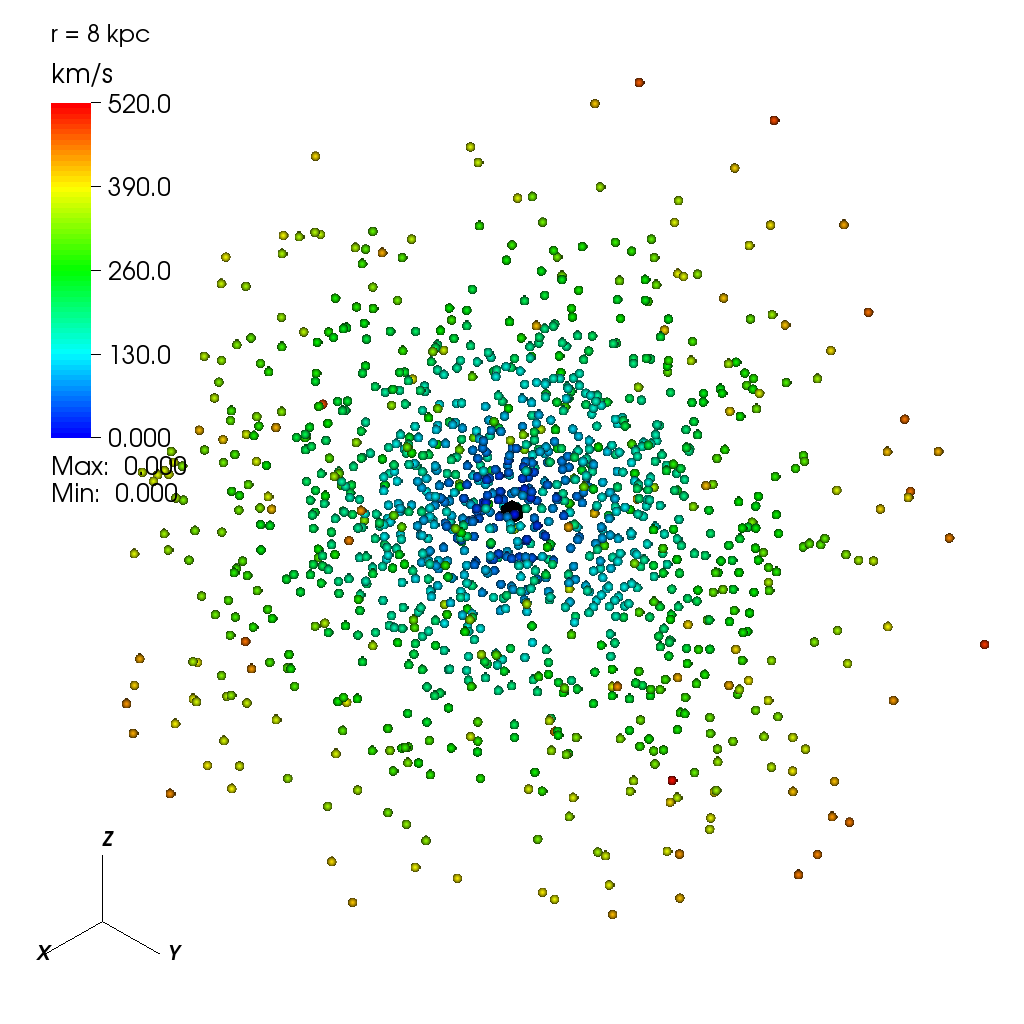}
	\includegraphics[width=0.315\textwidth,bb=0 0 1024 1024]{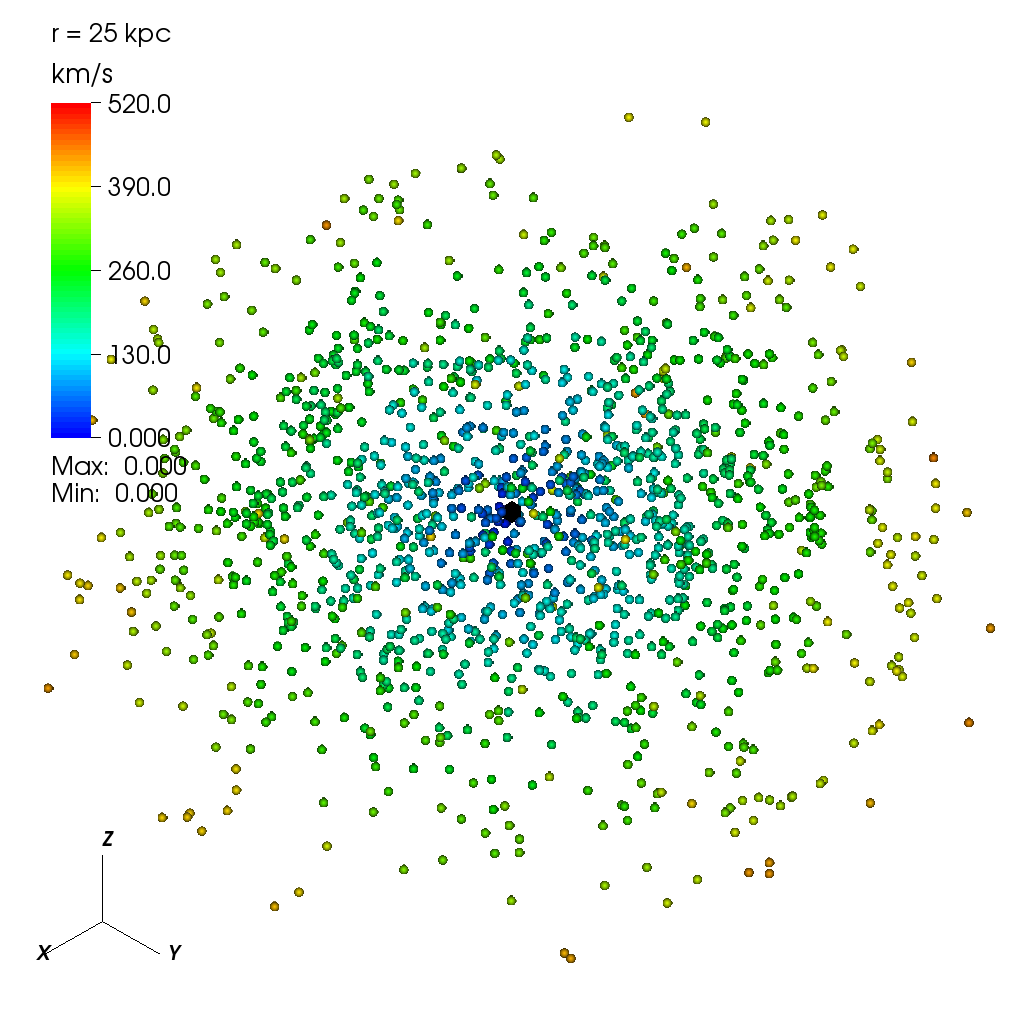}
	\includegraphics[width=0.315\textwidth,bb=0 0 1024 1024]{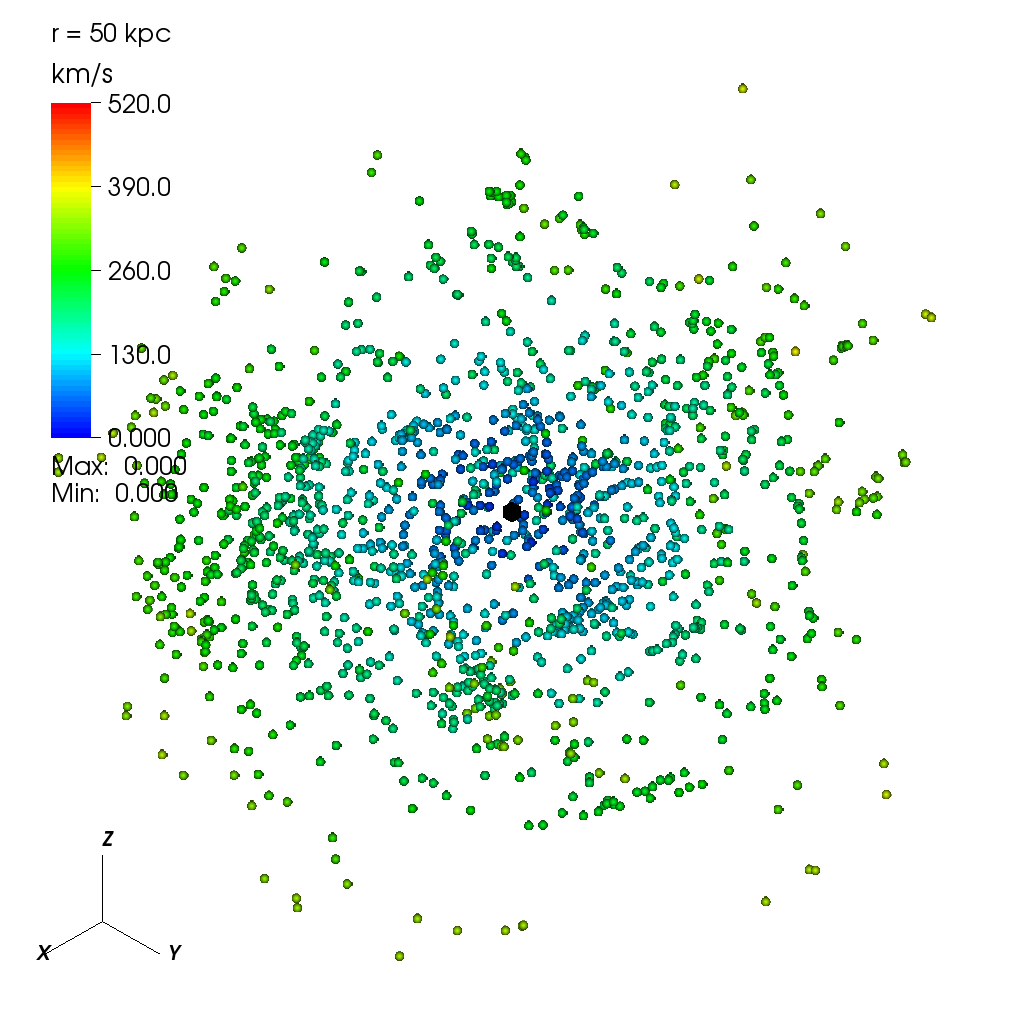}\\
	\includegraphics[width=0.315\textwidth,bb=0 0 1024 1024]{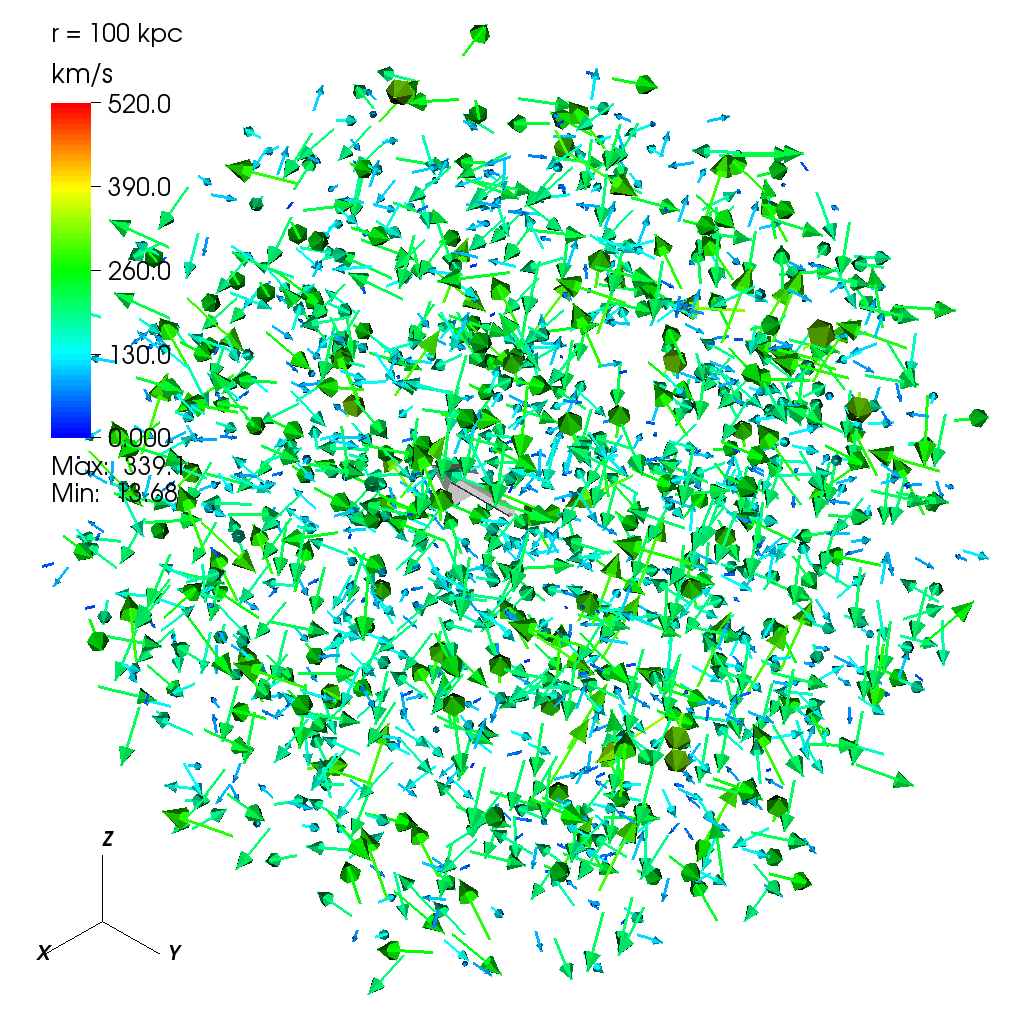}
	\includegraphics[width=0.315\textwidth,bb=0 0 1024 1024]{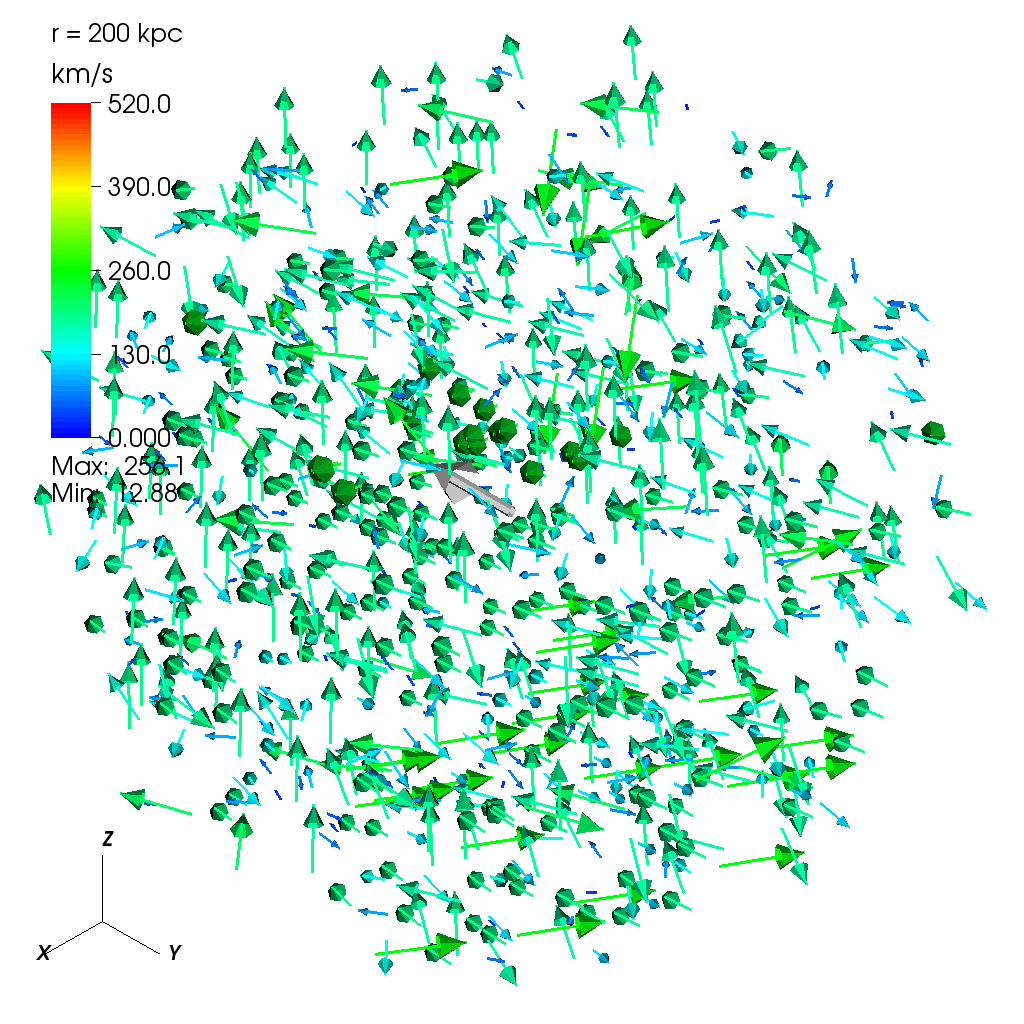}
	\includegraphics[width=0.315\textwidth,bb=0 0 1024 1024]{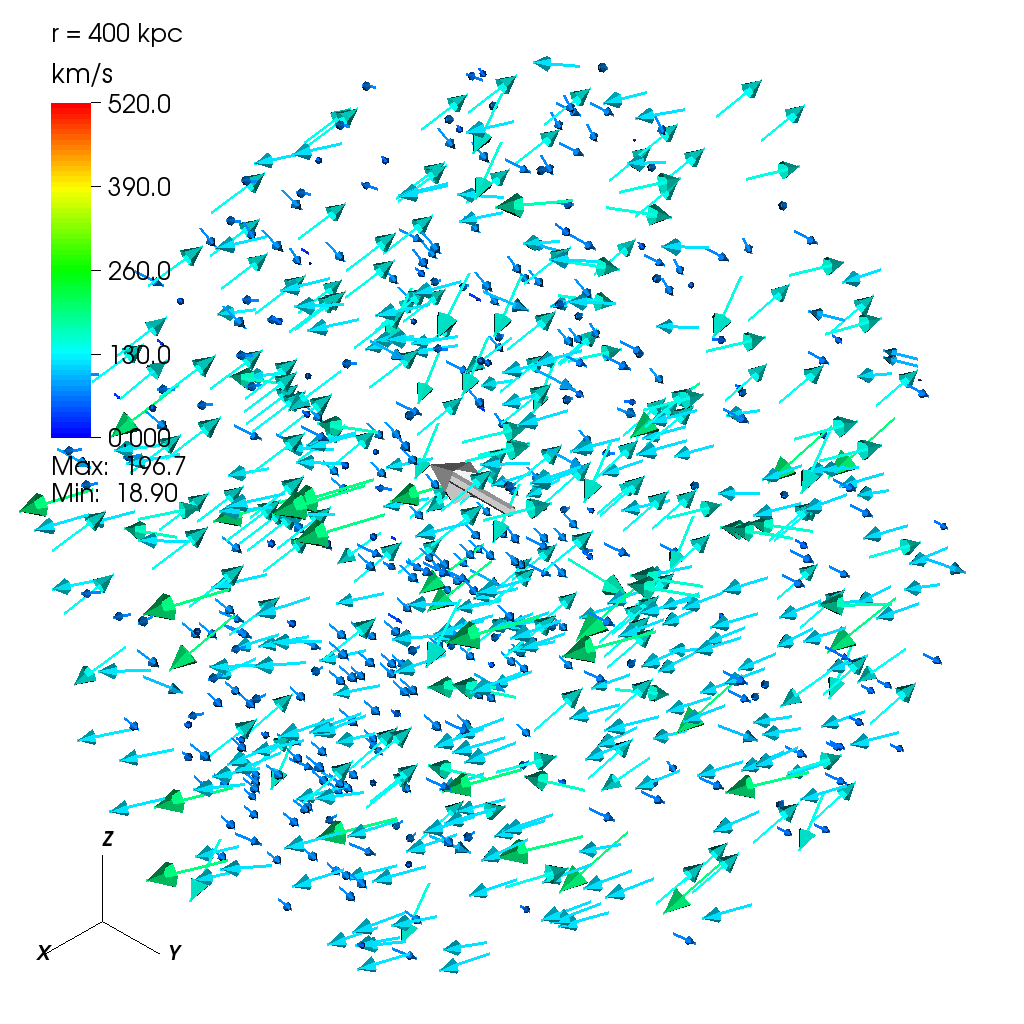}\\	
	\includegraphics[width=0.315\textwidth,bb=0 0 1024 1024]{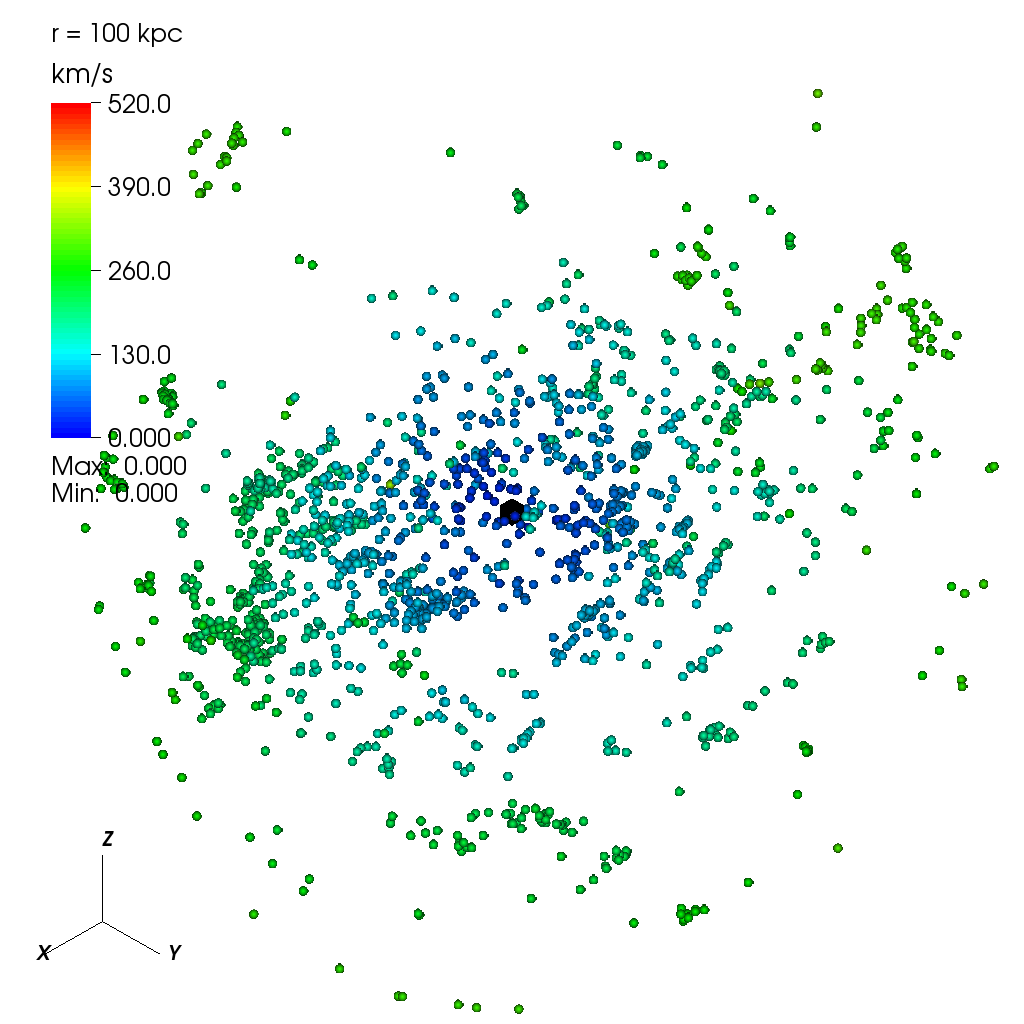}
	\includegraphics[width=0.315\textwidth,bb=0 0 1024 1024]{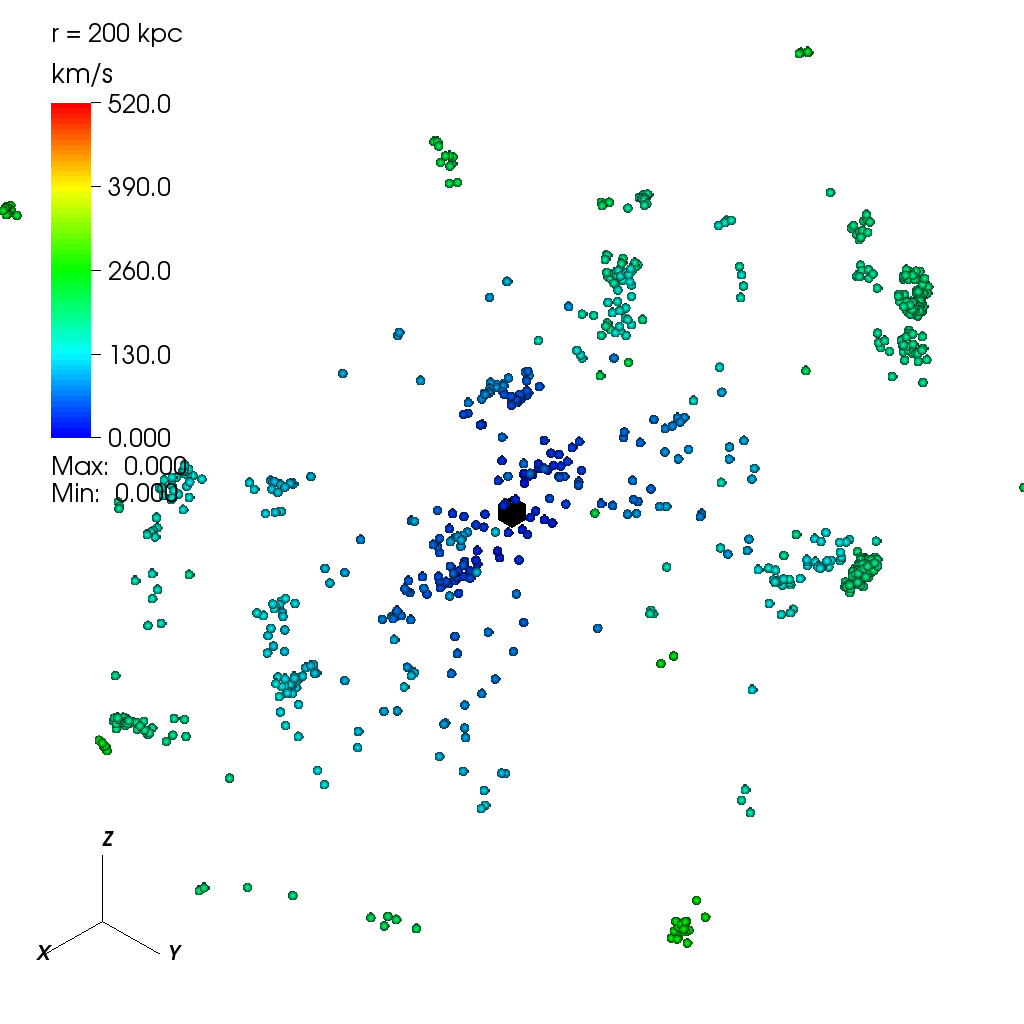}
	\includegraphics[width=0.315\textwidth,bb=0 0 1024 1024]{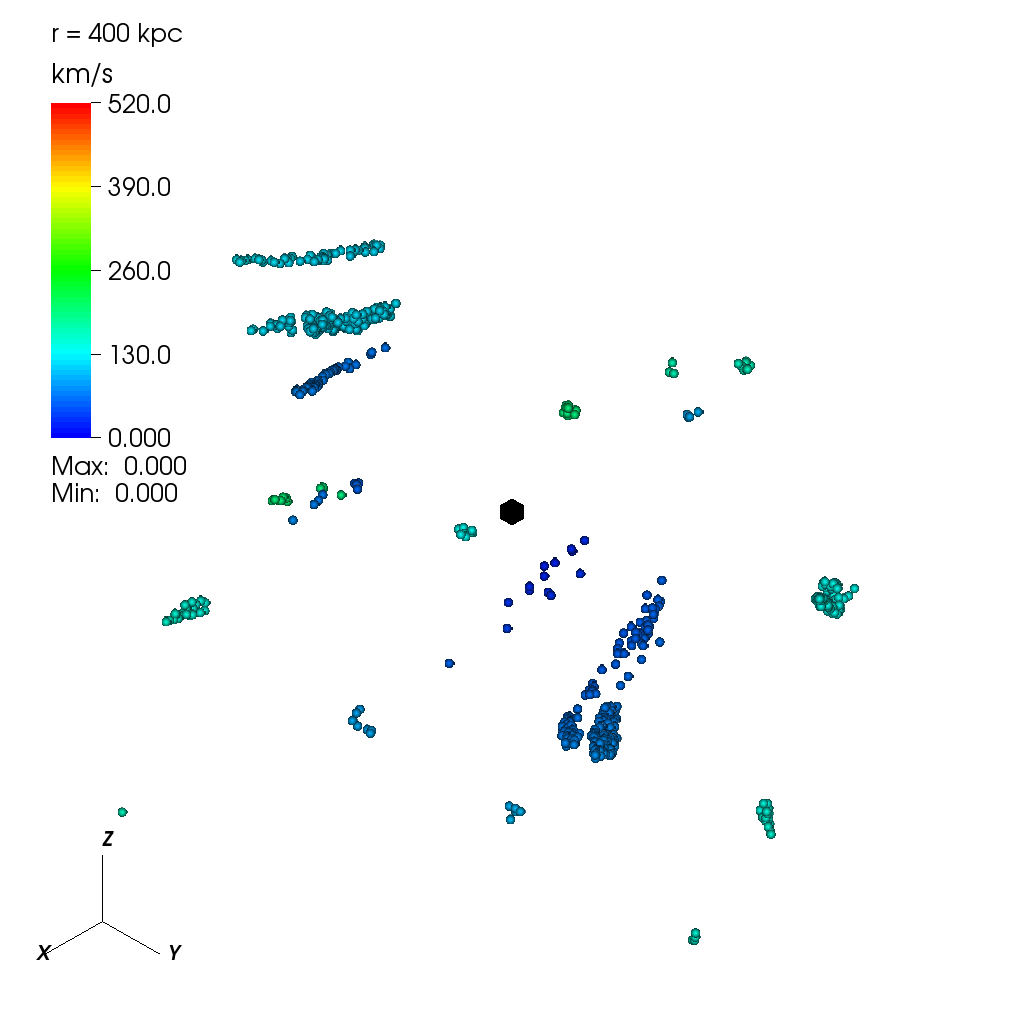}	
	\caption{Positions of particles within the selected spheres along the $y$-axis with velocity vector attached (1st and 3rd row) as well as the corresponding location of the particles in the local velocity space (2nd and 4th row). The colour in both cases is given by the magnitude of the velocity vector.}
	\label{fig:velstruc}
\end{figure*}

\begin{figure*}
	\centering
	\includegraphics[width=0.495\textwidth,bb=0 0 1024 1024]{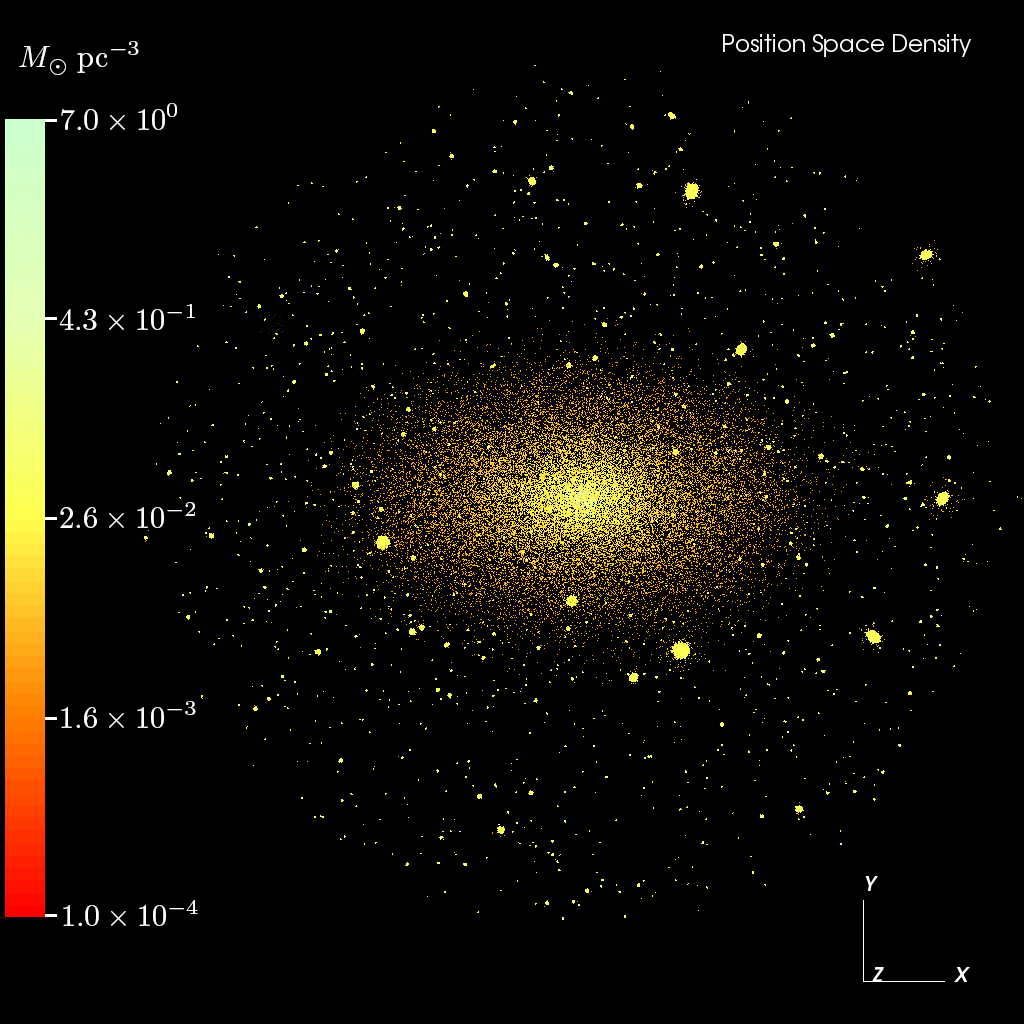}
	\includegraphics[width=0.495\textwidth,bb=0 0 1024 1024]{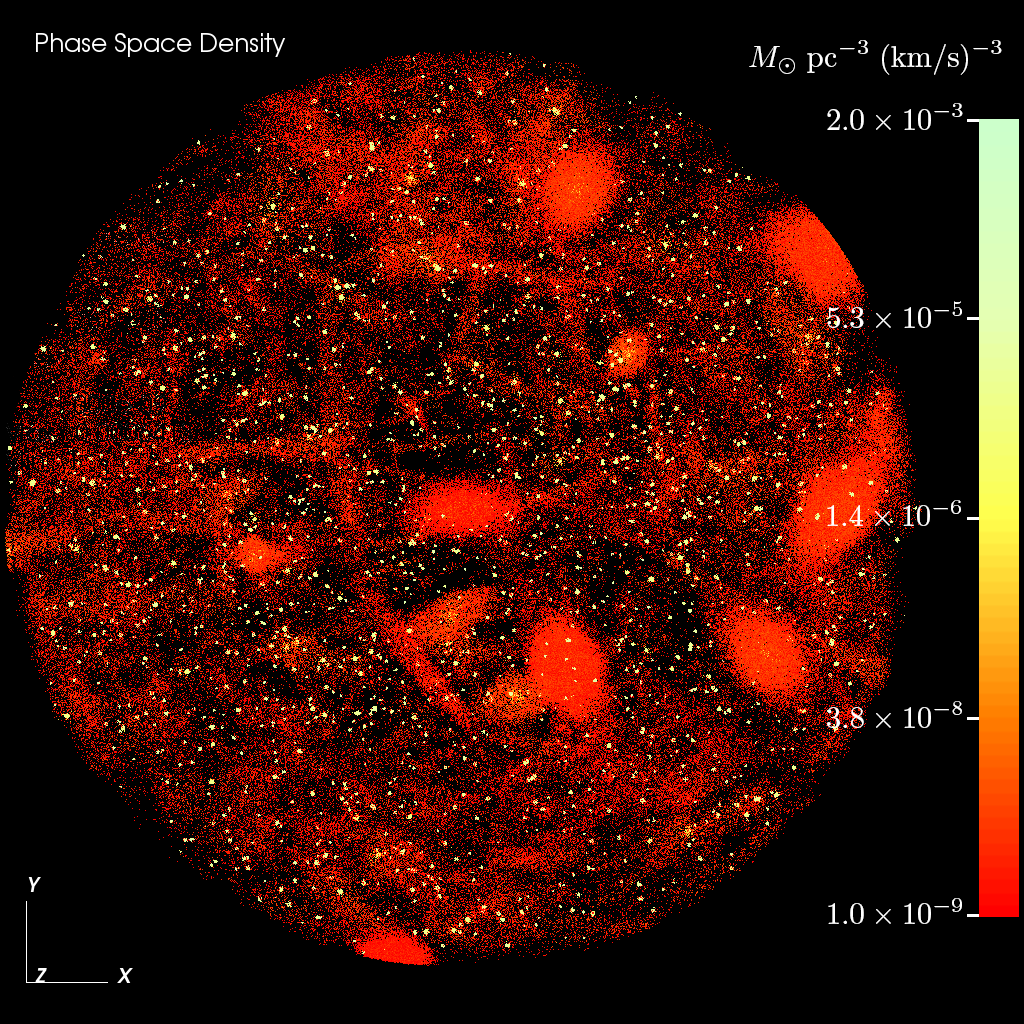}
	\caption{Central sphere of radius 50 kpc where we calculated the position space density (left) and the true phase space density (right) with EnBiD. Approximately 2000 peaks from the subhaloes within that radius are visible in both pictures but the contrast of subhaloes is higher in the phase space density picture. In the phase space picture are also large scale dark matter streams visible which were formed by tidal mass loss of infalling subhaloes. These streams are not visible in the traditional position space density picture.}
	\label{fig:denpsd}
\end{figure*}

We now turn to a closer look at the local velocity space structure. For this purpose, we present 3D visualizations of the velocity structure within spheres cut out along the $y$-axis, with radii given by $r_{\mathrm{sph}}(r)$ from Table \ref{tab:shellsummary}.

Fig. \ref{fig:velstruc} shows the positions and velocity vectors for every particle within these spheres (1st and 3rd row) for all six galactocentric distances we investigated. We also plot the location of the particles in the local velocity space (2nd and 4th row). The colour (and length in position space) encodes the magnitude of the velocity vector. The big white arrow in the centre of the position space plots points towards the galactic centre whereas the black cube in the centre of the local velocity space plots marks the origin. For the velocity space plots, we split the velocity vector field in a radial- ($x$-axis), $\varphi$- ($y$-axis) and a $\vartheta$-component ($z$-axis).

These spheres have been selected to be free of subhaloes. Therefore, one would not expect to see clumpy structure in the phase space of these spheres. In the inner spheres no velocity space structure is apparent. The orientation of the velocity vectors appears random and the actual velocity distribution can be well fit by a multivariate normal distribution (the different velocity dispersion components are given in Table \ref{tab:shellsummary}). But further out, there is evidence for locally coherent motion, visible as groups of vectors with the same colour pointing in the same direction. This is even clearer in the velocity space plots, which exhibit an increasing degree of clumpiness the further out one goes. At 400 kpc, for example, there are only a bit more than a dozen clumps in velocity space and no smooth component at all. Obviously a smooth multivariate normal distribution would not provide a good fit. This is evidence that the outer part of the halo is built up from a collection of large scale streams from the tidal disruption of infalling subhaloes. But so too is the smooth component in centre, in that it also likely consists of an overlap of many, many streams. The central limit theorem then guarantees that the resulting distribution in the centre closely resembles something like a multivariate normal distribution \citep{2003MNRAS.339..834H} or a generalisation thereof \citep{2006JCAP...01..014H}.

In Fig. \ref{fig:denpsd} we show the central region of Via Lactea II: the sphere has a radius of 50 kpc. We plot the position space density (left) and the true phase space density (right) calculated with EnBiD\footnote{\texttt{http://sourceforge.net/projects/enbid/}} \citep{2006MNRAS.373.1293S}. It is important here to calculate the true phase space density in 6D and not the commonly used pseudo phase space density $\rho/\sigma^3$ (where $\sigma$ is the velocity dispersion), since in the latter information in velocity space is lost by averaging over the particles in local position space, instead of calculating the density in local velocity space directly from the particles. We only show the top 5 to 6 orders of magnitude in position and phase space density, in order not to overload the two pictures.

Many cold streams are clearly visible in the central region. Although these cold streams only contribute a few percent to the local mass density, their velocity dispersion is just a few $\km~\s^{-1}$, resulting in a very high phase space density for these particles. These streams are not visible in traditional density or density squared pictures and can only be revealed by visualising the true phase space density. Via Lactea II seems to be the first structure formation simulation with sufficient resolution to reveal these streams in phase space, since previous lower resolution runs did not show such phase space features. Also subhaloes are better visible due to their higher contrast in phase space density. Approximately 2000 peaks are seen in the central 50 kpc sphere phase space density image. The high contrast makes the phase space density the ideal method of finding subhaloes.

From our findings it is obvious that also the central region shows a huge amount of additional structure aside from the expected subhalo density peaks. A more detailed discussion about streams and their properties will follow in a future publication.

\section{Discussion and conclusions}\label{sec:conclusions}

In this paper we present a study of local properties of the Via Lactea II dark matter halo. Commonly, characteristics of dark matter haloes are described by spherically symmetric profiles. Here, we show that locally, at a fixed galactocentric distance, these properties can vary by orders of magnitude from their canonical, spherically averaged values. For example, while the density is smooth in the centre, in the outskirts the density range spans four orders of magnitude. This is due to the presence of both subhaloes and underdense regions or holes in the matter distribution. The widespread assumption that there are no bulk flows is only warranted in the central region, where we also find the local velocity dispersion ellipsoid to be aligned with the shape ellipsoid of the halo. Neither of these findings hold in the outer parts of the halo, where particles exhibit bulk motions and a more isotropic distribution of the velocity dispersion ellipsoid's orientation is found. The local velocity space structure can, in general, not be well described by a smooth multivariate normal distribution, at least not in the outer parts of the halo. A qualitatively new feature in Via Lactea II is the detection of streams that are made visible only through their true 6D phase space density.

Such variations of local properties is due to both the triaxial shape of the dark matter halo and its generally clumpy structure in phase space. We find that the phase space structure of a dark matter halo shows a significant departure from the canonical picture of a smooth background density profile with subhaloes in position space and multivariate normal distributions in velocity space. Spherically averaged quantities do not adequately describe properties of the dark matter halo at a given galactocentric distance - especially not the wealth of structure we find locally - since a lot of information gets smeared out by spherically averaging. In general, dark matter haloes show a high degree of graininess - clumps in phase space - which will probably show up at even smaller scales that now seem to be smooth in future simulations with higher resolution. Therefore, a smooth and featureless distribution function does not accurately describe dark matter haloes that form in a cosmological structure formation simulation.

We find several correlations between the shape ellipsoid and the local velocity dispersion ellipsoid. It is unclear at the moment to what degree our findings are universal or how these correlations depend on the hierarchical build up history.

\citet{2006PASA...23..125K} earlier estimated analytically the effect of a triaxial halo shape on the variance of the local density. Their values of the dispersion normalised to the spherically averaged value for a halo with a similar triaxiality correspond quite good with our values in the centre of the halo but we get much higher dispersion values (larger than the mean) in the outskirts of the halo. This difference is due to their assumption of a smooth triaxial density profile and we get a higher scatter due to the presence of subhaloes and underdense regions.

For dark matter direct detection experiments it is essential to know the local dark matter properties at 8 kpc. The spherically averaged value for the density in Via Lactea II is $\tilde{\rho}(8~\kpc) = 1.056 \times 10^{-2} ~\Mo~\pc^{-3} = 0.4008~\GeV~c^{-2}~\cm^{-3}$, close to $0.3~\GeV~c^{-2}~\cm^{-3}$ which is often used as a canonical value in the literature \citep{1998PhRvD..57.3256K,2008PhLB..667..212P}. But as we have shown here, this value can vary locally. One of the large uncertainty factors is the missing information about the orientation of the disk with respect to the dark matter halo. There are different claims from observations and theory about a possible alignment of the angular momentum axis of the disk with the shape axes of the halo \citep[see e.g.][and references therein]{2004ApJ...613L..41N,2005ApJ...627L..17B,2005ApJ...628...21S} so that a clear answer is not possible at the moment. But often it is claimed that the angular momentum axis of the disk and the short axis ($z$) tend to be aligned \citep{2005ApJ...627L..17B} which would mean that the disk would lie preferentially in the $xy$-plane in our coordinate system. An additional problem is that we measure these local properties within a sphere of $r_{\mathrm{sph}}(8~\kpc) = 500~\pc$ radius but for dark matter detection experiments more a scale of $1~\AU = 4.848 \times 10^{-6}~\pc = 9.696 \times 10^{-9}~r_{\mathrm{sph}}(8~\kpc)$ is relevant and it is not clear what the local properties of the dark matter distribution on a 1 AU scale are or how one could reasonably extrapolate that over 8 orders of magnitude - especially considering the highly non-linear numerical effects mentioned above and in the appendix that affect the local phase space structure. Also the missing baryonic physics in Via Lactea II like adiabatic contraction, stellar disk and bulge, inspiralling compact objects like black holes etc. can modify the central dark matter structure in either way. Therefore, it is still not clear what the detailed structure of the dark matter locally is.

\section*{Acknowledgement}

The Via Lactea II simulation was carried out at the Oak Ridge National Laboratory through an award from the Department of Energy's Office of Science as part of the 2007 Innovative and Novel Computational Impact on Theory and Experiment (INCITE) program\footnote{\texttt{http://www.science.doe.gov}}. It is especially a pleasure to thank Bronson Messer and the Scientific Computing Group at the National Center for Computational Sciences\footnote{\texttt{http://www.nccs.gov}} for their support and help. It is also a pleasure to thank Glenn van de Ven for helpful comments. Further supporting computations for generating the initial conditions, code optimisation and analysis were performed on MareNostrum at the Barcelona Supercomputing Centre\footnote{\texttt{http://www.bsc.org.es}}, the zBox at University of Zurich, Columbia at NASA Ames\footnote{\texttt{http://www.nas.nasa.gov}} and on Pleiades at University of California Santa Cruz\footnote{\texttt{http://pleione.ucsc.edu/pleiades}}. M. Z. gratefully acknowledges support by the Swiss National Science Foundation. Additionally, support for this work was provided by NASA through grants HST-AR-11268.01-A1 and NNX08AV68G (P. M.) and Hubble Fellowship grant HST-HF-01194.01 (J. D.). M. K. acknowledges support from the William L. Loughlin Fellowship at the Institute for Advanced Study.

\bibliography{RDB_S}

\appendix

\section{Comparisons}\label{sec:comparison}

\begin{figure*}
	\centering
	\ifthenelse{\boolean{useepsfigures}}{
	\includegraphics[width=0.495\textwidth]{shell8kpc_rhocomp}
	\includegraphics[width=0.495\textwidth]{shell25kpc_rhocomp}\\
	\includegraphics[width=0.495\textwidth]{shell100kpc_rhocomp}
	\includegraphics[width=0.495\textwidth]{shell400kpc_rhocomp}
	}{
	\includegraphics[width=0.495\textwidth,bb=0 0 574 574]{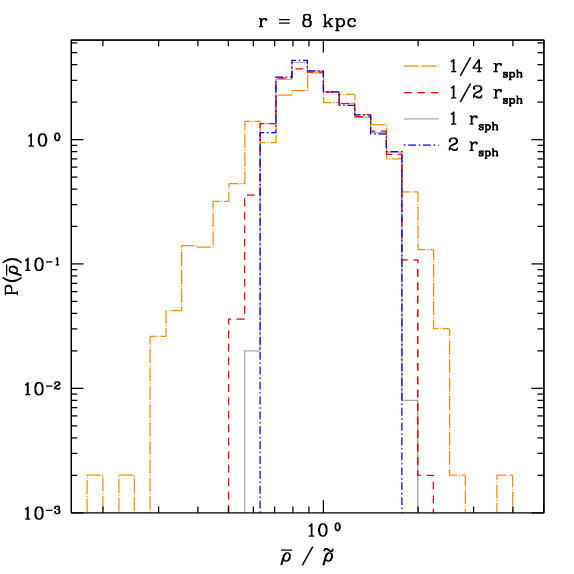}
	\includegraphics[width=0.495\textwidth,bb=0 0 574 574]{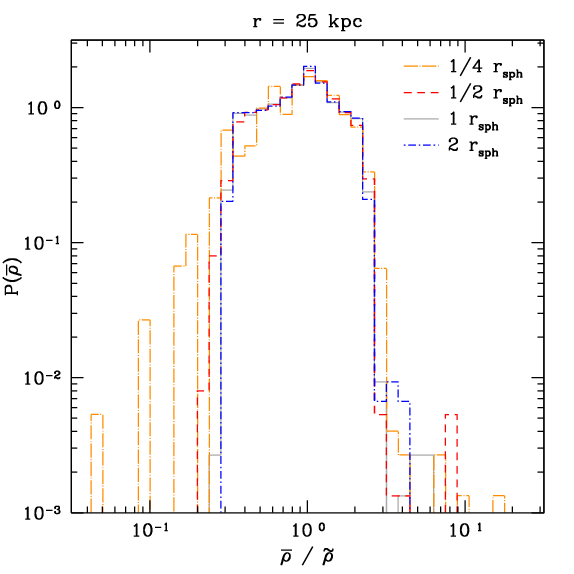}\\
	\includegraphics[width=0.495\textwidth,bb=0 0 574 574]{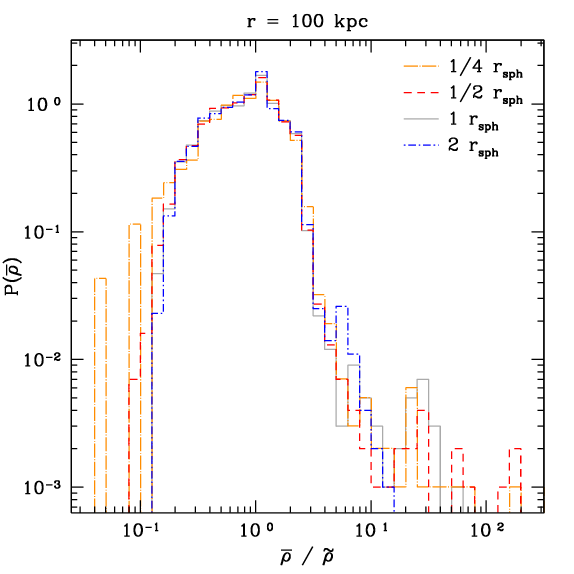}
	\includegraphics[width=0.495\textwidth,bb=0 0 574 574]{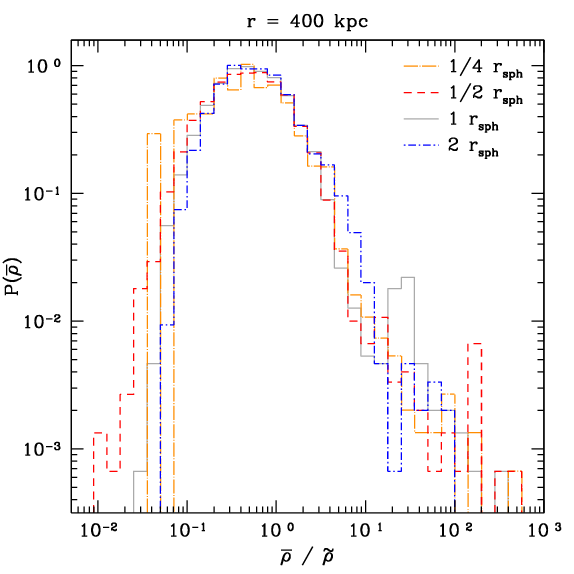}
	}
	\caption{Probability density functions of the local density $\bar{\rho}$ at different galactocentric distances $r$ for different definitions of locality normalised by the standard spherically averaged value $\tilde{\rho}$ given in Table \ref{tab:shellsummary}. An average density sphere of radius 1 $r_{\mathrm{sph}}$ contains 1356 particles. The 1/4 $r_{\mathrm{sph}}$ probability density function is significantly broadened by Poisson noise.}
	\label{fig:rhocomp}
\end{figure*}

\begin{figure*}
	\centering
	\ifthenelse{\boolean{useepsfigures}}{
	\includegraphics[width=0.495\textwidth]{shell8kpc_rhocomp2}
	\includegraphics[width=0.495\textwidth]{shell25kpc_rhocomp2}\\
	\includegraphics[width=0.495\textwidth]{shell100kpc_rhocomp2}
	\includegraphics[width=0.495\textwidth]{shell400kpc_rhocomp2}
	}{
	\includegraphics[width=0.495\textwidth,bb=0 0 574 574]{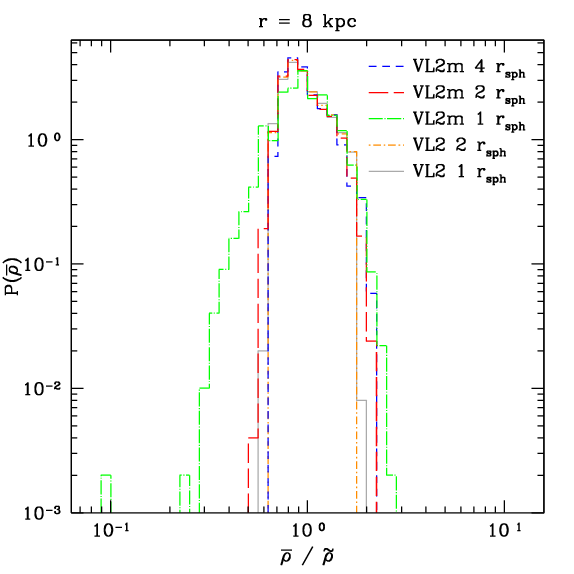}
	\includegraphics[width=0.495\textwidth,bb=0 0 574 574]{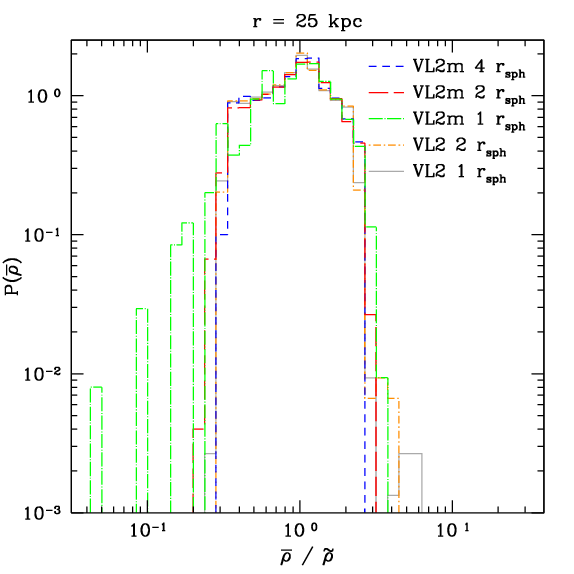}\\
	\includegraphics[width=0.495\textwidth,bb=0 0 574 574]{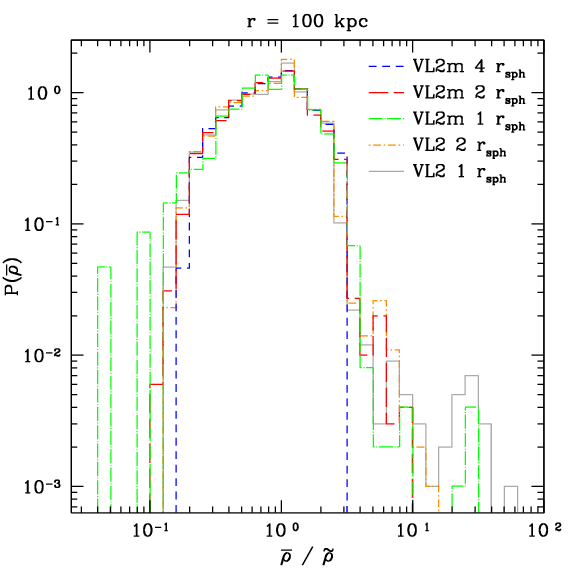}
	\includegraphics[width=0.495\textwidth,bb=0 0 574 574]{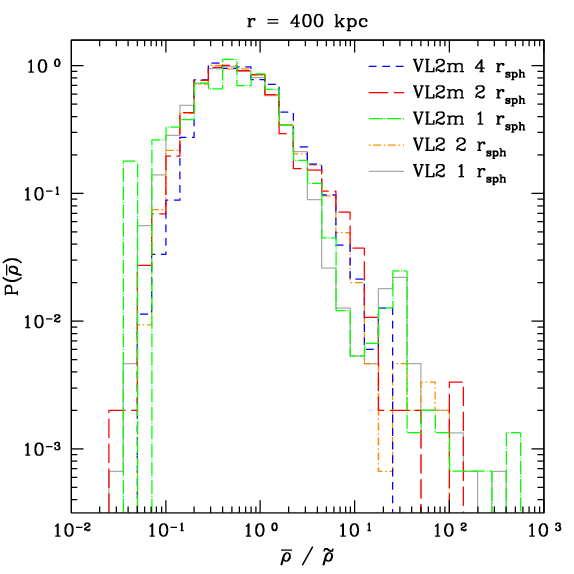}
	}
	\caption{Probability density functions of the local density $\bar{\rho}$ for different resolutions of Via Lactea II normalised by the  spherically averaged value $\tilde{\rho}$ for the high resolution run given in Table \ref{tab:shellsummary}. An average density sphere of radius 1 $r_{\mathrm{sph}}$ contains 1356 
particles in VL2, and 21 particles in VL2m. This means the 1 $r_{\mathrm{sph}}$ measurements in the VL2m simulation are rather noisy.}
	\label{fig:rhocomp2}
\end{figure*}

In this section we investigate the influence of different definitions of locality and numerical resolution on our local estimates.

\subsection{Definition of locality}

Our definition of locality is somewhat arbitrary. If the region is too big then local properties get smeared out by the averaging procedure and if the region is too small then statistical fluctuations become important. Hence, we chose the size of our spheres so that they contain $\mathcal{O}(10^3)$ particles as in similar previous work \citep{2001PhRvD..64f3508M,2002PhRvD..66f3502H}. Exactly: an average density sphere of radius 1 $r_{\mathrm{sph}}$ contains 1356 particles. In order to see the effect of different choices for our local estimate, we repeated the exercise from section \ref{sec:properties} with spheres of different radii of the following size: $1/4~r_{\mathrm{sph}}$, $1/2~r_{\mathrm{sph}}$, and $2~r_{\mathrm{sph}}$, with a corresponding increase or decrease in the number of particles sampled per sphere. The resulting Poisson scatter ranges from $\sim$22\% for the smallest spheres ($1/4~r_{\mathrm{sph}}$) to $\sim$1\% for the largest spheres ($2~r_{\mathrm{sph}}$) if we take the spherical averaged density as a reference. By only looking at the lowest density along the $z$-axis we get a range of $\sim$41\% for the smallest spheres to $\sim$1.7\% for the largest spheres.

In Fig. \ref{fig:rhocomp}, we present the effect of the different definitions of locality on the local density distribution and see that they show the expected result. The regions that contain 8 times more particles are of course smoothed to a higher degree so that the resulting spread in the probability distribution function is a bit reduced. In a similar way the spread for the 8 times smaller spheres is increased. But the different definitions of locality mainly affect the rare outliers on the tails of the distribution. By reducing to 64 times less particles, the extremes of the distribution become much more populated, which is mostly due to the increased Poisson scatter and to a lesser extend due to increased graininess on the smaller probed scales.

Any size of a sphere that contains at least a few hundred particles would be fine for a definition of locality, indicating that our choice is a rather conservative choice and the resulting variances of the different distributions from the previous sections are lower limits. The probability density functions for the other characteristics show the same effects as function of locality definition as the one discussed here for the density.

\subsection{Influence of resolution}

We compare the results from Via Lactea II (VL2) to a medium resolution simulation (VL2m) of the same halo. In the medium resolution run, the particle mass was a factor 64 higher than in the high resolution run. Therefore, in order to probe local properties with approximately the same number of particles spheres with $\sqrt[3]{64} = 4$ times larger radii were used.

In Fig. \ref{fig:rhocomp2} we show the probability density functions of the local density for the medium and high resolution run for different sizes of the spheres. In general the rarer peaks at the low and high end of the distribution are missing in the medium resolution run when one compares spheres with equal number of particles in both simulations (i.e. VL2 with 1 $r_{\mathrm{sph}}$ and VL2m with 4 $r_{\mathrm{sph}}$) since the low resolution run resolves less substructure and it is smoothed over a 4 times larger scale. When comparing spheres of equal physical size in both runs, then the low resolution run shows a broader distribution due to additional Poisson noise.

The lack of high density peaks in the low resolution simulation is due to numerical effects. Less subhaloes survive in the medium resolution run since the subhaloes are resolved with fewer particles and are more easily tidally disrupted. This leads to the effect that a larger region in the centre is smooth in lower resolution runs. The higher the resolution the smaller this apparently smooth region is.

Additionally, the higher degree of numerical artefacts (e.g. heating, relaxation etc.) in the medium resolution run smears out streams. For example, no dark matter streams are visible in the inner 50 kpc in phase space density maps of the medium resolution of Via Lactea II and the peak phase space densities of the subhaloes are much lower than the values from the high resolution run. 

It is not clear at the moment to what degree the dark matter streams in the high resolution run are broadened by artificial heating in the simulations. Of course there are also real dynamical heating sources, such as dark matter subhaloes or baryonic structures like a stellar disk, a bulge, or gas clouds \citep[see e.g.][]{2002MNRAS.332..915I,2008ApJ...681...40S}.

Material that ends up in the central regions of dark matter haloes originates from early forming halos \citep{2005MNRAS.364..367D} and was accreted into the main halo early on \citep{2002PhRvD..66f3502H}. One therefore expects a higher degree of phase mixing due to the early accretion and short time-scales in the centre. Nevertheless it is not clear if the smooth appearance of the central regions in Via Lactea II is entirely due to effcient phase mixing: The small, early progenitors that build up the central dark matter halo of Via Lactea II are under resolved and therefore too low numerical resolution might appear as an efficient phase mixing.

\section{Balls-in-bins statistics}\label{sec:bib}

Let's try to solve the following problem. We have $n$ bins and balls are thrown randomly into the bins. What is the probability $p(k,l)$ of finding bins that contain $k$ balls after throwing $l$ balls?

The probability of a ball hitting a bin is given by $p_h = 1/n$ and the probability of missing a bin is given by $p_m = 1 - p_h = 1 - 1/n$. Obviously, we have $p(0,0) = 1$, $p(k,0) = 0$ for $k>0$ and $p(k,l) = 0$ for $k>l$. After throwing $l-1$ balls we can write
\begin{equation}
p(k,l) = p_m p(k,l-1) + p_h p(k-1,l-1)~.
\end{equation}
This is a recursion formula that allows us now to construct the general form of $p(k,l)$. By setting $p(-1,l) = 0$ for all $l$, we get for the non-zero values for the first few values of $l$
\begin{eqnarray}
p(0,1) & = & p_m\\
p(1,1) & = & p_h\\
p(0,2) & = & p_m^2\\
p(1,2) & = & 2 p_m p_h\\
p(2,2) & = & p_h^2\\
p(0,3) & = & p_m^3\\
p(1,3) & = & 3 p_m^2 p_h\\
p(2,3) & = & 3 p_m p_h^2\\
p(3,3) & = & p_h^3
\end{eqnarray}
and so on. It becomes obvious that the general pattern is given by a binomial distribution
\begin{eqnarray}
p(k,l) & = & \left(\! {\begin{array}{c} l \\ k \\ \end{array}} \!\right) p_m^{l-k} p_h^{k} \\
& = & \left(\! {\begin{array}{c} l \\ k \\ \end{array}} \!\right) \left( 1-\frac{1}{n} \right)^{l-k} \left( \frac{1}{n} \right)^{k}~.
\end{eqnarray}

We are specifically interested in the case of getting empty bins (or spheres in our case), i.e. $k = 0$. This results in
\begin{equation}
p_{\mathrm{empty}}(\lambda) \equiv p(0,l) = \left( 1-\frac{1}{n} \right)^{l} = \left( 1-\frac{1}{n} \right)^{n \lambda}
\end{equation}
where $\lambda = l/n$ is the expectation value of balls per bin. We can simplify this in the large $n$ limit to
\begin{equation}
p_{\mathrm{empty}}(\lambda) = \lim_{n\rightarrow\infty} \left[\left( 1-\frac{1}{n} \right)^{n} \right]^\lambda = \mathrm{e}^{-\lambda}
\end{equation}
which we use to estimate the expected fraction of empty bins respectively spheres.

\label{lastpage}

\end{document}